\documentclass{aa}

\usepackage{txfonts}
\usepackage{graphicx}
\usepackage{float}
\usepackage{amsmath}
\usepackage{epsfig}
\usepackage{color}
\usepackage{array}
\usepackage[pdftex,colorlinks=true,allcolors=blue,pdfstartview=Fit,
  bookmarks=true,bookmarksnumbered=true]{hyperref}

\newcommand{\ve}[1]{\mathbf{#1}}

\newcommand{\Nobs}{N_\text{obs}}
\newcommand{\Ncomp}{N_\text{comp}}
\newcommand{\Nmix}{N_\text{mix}}

\newcommand{\A}{\ve{A}}
\newcommand{\G}{\ve{G}}
\renewcommand{\u}{\ve{u}}
\newcommand{\U}{\ve{U}}

\newcommand{\T}{\ve{T}}
\newcommand{\Q}{\ve{Q}}
\newcommand{\WQ}{\widetilde{\Q}}
\newcommand{\D}{\ve{D}}
\newcommand{\F}{\ve{F}}
\newcommand{\Z}{\ve{Z}}

\renewcommand{\S}{\ve{S}}
\newcommand{\lmax}{{\ell_\text{max}}}

\newcommand{\N}{\ve{N}}
\newcommand{\M}{\ve{M}}
\newcommand{\R}{\ve{R}}
\newcommand{\B}{\ve{B}}
\newcommand{\Y}{\ve{Y}}

\renewcommand{\P}{\ve{P}}
\renewcommand{\d}{\ve{d}}
\newcommand{\x}{\ve{x}}
\newcommand{\e}{\ve{e}}
\renewcommand{\r}{\ve{r}}
\renewcommand{\b}{\ve{b}}
\newcommand{\s}{\ve{s}}
\newcommand{\n}{\ve{n}}
\newcommand{\W}{\ve{W}}
\newcommand{\I}{\ve{I}}

\newcommand{\mat}[1]{\left[ \begin{array}{ccccccccccccccc} #1 \end{array} \right]}

\newcommand{\Nside}{N_\text{side}}

\begin{document}

    \title{Multi-resolution Bayesian CMB component separation through Wiener-filtering with a pseudo-inverse preconditioner}

    \author{D. S. Seljebotn\inst{1}
            \and
            T. B{\ae}rland\inst{2}
            \and
            H. K. Eriksen\inst{1}
            \and
            K.-A. Mardal\inst{2}
            \and
            I. K. Wehus\inst{1}
        }

    \date{Received ?? / Accepted ??}

    \institute{%
      Institute of Theoretical Astrophysics, University of Oslo, P.O.\ Box 1029 Blindern, N-0315 Oslo, Norway \\
      \email{d.s.seljebotn@astro.uio.no}
      \and
      Department of Mathematics, University of Oslo, P.O.\ Box 1080 Blindern, N-0316 Oslo, Norway
    }
    \abstract{
      We present a Bayesian model for multi-resolution CMB component
      separation based on Wiener filtering
      and/or computation of constrained realizations, extending a
      previously developed framework.
      We also develop an efficient solver for the corresponding linear
      system for the associated signal amplitudes. The core of
      this new solver is an efficient preconditioner based on the
      \emph{pseudo-inverse} of the coefficient matrix of the linear system.
      In the full sky coverage case, the method gives
      a speed-up of 2--3x in compute time compared to a simple diagonal
      preconditioner, and it is easier to implement in terms of
      practical computer code.
      In the case where a mask is applied and prior-driven constrained realization is sought
      within the mask, this is the first time full convergence has been achieved
      at the full resolution of the Planck dataset.
      Prototype benchmark code is available at
      \url{https://github.com/dagss/cmbcr}.
    }

\keywords{methods: numerical -- methods: statistical --- cosmic microwave background}
\titlerunning{Multi-resolution CMB component separation using pseudo-inverse}
\maketitle

\section{Introduction}

The microwave sky has been observed on multiple frequencies by many
experiments, both ground-based, sub-orbital and satellite
experiments. Among the latter are the most recent NASA's WMAP mission
\citep{bennett:2013} and ESA's Planck mission
\citep{planck2015:mission}. The primary goal of all these experiments
is to produce the cleanest possible image of the true cosmological CMB
sky, and use this image to constrain the physics, contents and
evolution of the universe \citep[e.g.,][]{planck2015:parameters}. A
critical step in this process, however, is to remove the impact of
obscuring foreground emission from our own Galaxy, the Milky Way
\citep{planck2015:foregrounds}. This step is often called CMB
component separation, and informally refers to the process of
combining observations taken at different frequencies into a single
estimate of the true sky. Furthermore, as the sensitivity of new
generations of CMB experiments continue to improve, the importance of
CMB component separation is steadily increasing
\citep[e.g.,][]{bicep:2014}, and the field is today a central research
field in CMB cosmology, as the community has turned its attention to
detect sub-microkelvin gravitational waves.

One popular class of component separation methods for full sky CMB
estimation and component separation is through optimal Bayesian
analysis, in particular as implemented through the Gibbs sampling and
Wiener-filtering algorithms
\citep{gibbs-jewell,gibbs-wandelt,joint-bayesian-compsep}. This
approach, as implemented in the Commander computer code
\citep{gibbs-eriksen}, has been established by the Planck
collaboration as a standard method for astrophysical component
separation and CMB extraction \citep{planck2015:foregrounds}. However,
the Commander implementation that was used in the Planck 2015 (and
earlier) analysis suffers from one important limitation; it requires
all frequency channels to have the same effective instrumental beam,
or ``point-spread function''. In general, though, the angular
resolution of a given sky map is typically inversely proportional to
the wavelength, and the solution to this problem adopted by the Planck
2015 Commander analysis was simply to smooth all frequency maps to a
common (lowest) angular resolution. This, however, implies an
effective loss of both sensitivity and angular resolution, and is as
such highly undesirable. To establish an effective algorithm that
avoids this problem is the main goal of the current paper.

First, we start by presenting a native multi-resolution version of the
Bayesian CMB component separation model.  This is a straightforward
extension of the single-resolution model presented in
\cite{joint-bayesian-compsep}. Second, we construct a novel solver for
the resulting linear system, based on the \emph{pseudo-inverse} of the
corresponding coefficient matrix. We find that this solver
out-performs existing solvers in all situations we have applied it
to. Additionally, the full-sky version of the preconditioner is easier
to implement in terms of computer code than the simple diagonal
preconditioner that is most commonly used in the literature
\citep{gibbs-eriksen}. The big advantage, however, is most clearly
seen in the presence of partial sky coverage, where the speed-ups
reach factors of 100s. For full sky coverage, where a diagonal
preconditioner performs reasonably, the improvement is a more modest
2--3x.

The previous Commander implementation \citep{gibbs-eriksen} employs
the Conjugate Gradient (CG) method to solve the single-component CR
system, using a preconditioner based on computing matrix elements in
spherical harmonic domain. While this approach worked well enough for
WMAP, the convergence rate scales with the signal-to-noise ratio of
the experiment, and for the sensitivity of Planck, it becomes
impractical in the case of partial sky coverage, requiring several
thousand iterations for convergence, if it converges at all.  More
sophisticated solvers are described by
\cite{smith:2007,elsner-messaging}. We argue in \cite{multigrid} that
these appear to suffer from the same fundamental problem.  In
particular they quickly converge outside of the mask, but spend a long
time finding the solution inside the mask.

To our knowledge the multi-level, pixel-domain solver described by
\cite{multigrid} represents the state of the art prior to this
paper. However, the multi-level solver has some weaknesses which
became apparent in our attempt to generalize it for the purpose of
component separation. First, due to only working locally in pixel domain,
it is ineffective in deconvolving the signal on high $\ell$, where the
spherical harmonic transfer function of the instrumental beam falls
below $b_\ell=0.2$ or so. In the case of \cite{multigrid} this was not a problem
because the solution is dominated by a $\Lambda$CDM-type prior on the CMB
before reaching these scales. However, for the purposes of component separation
one wants to apply no prior or rather weak priors, and so the solver in \cite{multigrid}
fails to converge. The second problem is that it is challenging to work
in pixel domain with multi-resolution data where the beam sizes vary with
a factor of 10, as is the case in Planck. Thirdly, extensive tuning was required
on each level to avoid ringing problems. Finally, the algorithm in
\cite{multigrid} requires extensive memory-consuming precomputations.

In contrast, the solver developed in this
paper 1) offers very cheap precomputations; 2) does not depend on applying
a prior; 3) is robust with respect to the choice of statistical priors on the
component amplitudes; and 4) has much less need for tuning. The solver
combines a number of techniques, and fundamentally consists of two parts:
\begin{enumerate}
\item A \emph{pseudo-inverse preconditioner}, developed in
  Sect. \ref{sec:pseudo-inverse-preconditioner} and Appendix
  \ref{appendix:pseudo-inverse}, which improves on the simple diagonal
  preconditioner by better incorporating inhomogeneities in the RMS
  maps of the data. This performs very well on its own in cases
  with full sky coverage.
\item A \emph{mask-restricted solver} which should be applied in
  addition in cases of partial sky coverage. This solver is developed
  in Sect. \ref{sec:mask}.  By solving the system under the mask as a
  separate sub-step, we converge in about 20 CG iterations, rather
  than 1000s of iterations.
\end{enumerate}
These two techniques are somewhat independent, and it would be
possible to use the mask-restricted solver together with the older
diagonal preconditioner. Still we present them together in this paper,
in order to establish what we believe is the new state of the art for
solving the CMB component separation and/or constrained realization
problem.

\setcounter{footnote}{0}

\section{Bayesian multi-resolution CMB component separation}
\subsection{Spherical harmonic transforms in linear algebra}
We will assume that the reader is familiar with spherical harmonic
transforms (SHTs), and simply note that they are the spherical analogue of
Fourier transforms \citep[for further details see, e.g.,][]{libsharp}.
The present work requires us to be very specific about the properties
of these transforms as part of linear systems. We will write $\Y$ for
the transform from harmonic domain to pixel domain (\emph{spherical
  harmonic synthesis}), and $\Y^T \W$ for the opposite transform
(\emph{spherical harmonic analysis}).  The $\W$ matrix is diagonal,
containing the per-pixel quadrature weights employed in the analysis
integral for a particular grid.

Of particular importance for the present work is that there exists no perfect grid
on the sphere; all spherical grids have slightly varying distances between
grid points. This means that some parts of the sphere will see smaller scales
than other parts, and that, ultimately, there is no \emph{discrete} version
of the spherical harmonic transform analogous to the discrete Fourier transforms
which maps $\mathbb{R}^n$ to $\mathbb{R}^n$. Specifically, $\Y$ is rectangular
and thus not invertible. Usually $\Y$ is configured
such that the number of rows (pixels) is greater than the number of
columns (spherical harmonic coefficients), in which case an harmonic
signal is unaltered by a round-trip to pixel domain so that
$\Y^T \W \Y = \I$. In that case, the converse operator, $\Y \Y^T \W$, is singular,
but in a very specific way: It takes a pixel map and removes any scales from it
that are above the band-limit $L$.

\subsection{The component separation model}
\label{sec:model}

\begin{figure}
  \centering
  \includegraphics[width=0.8\linewidth]{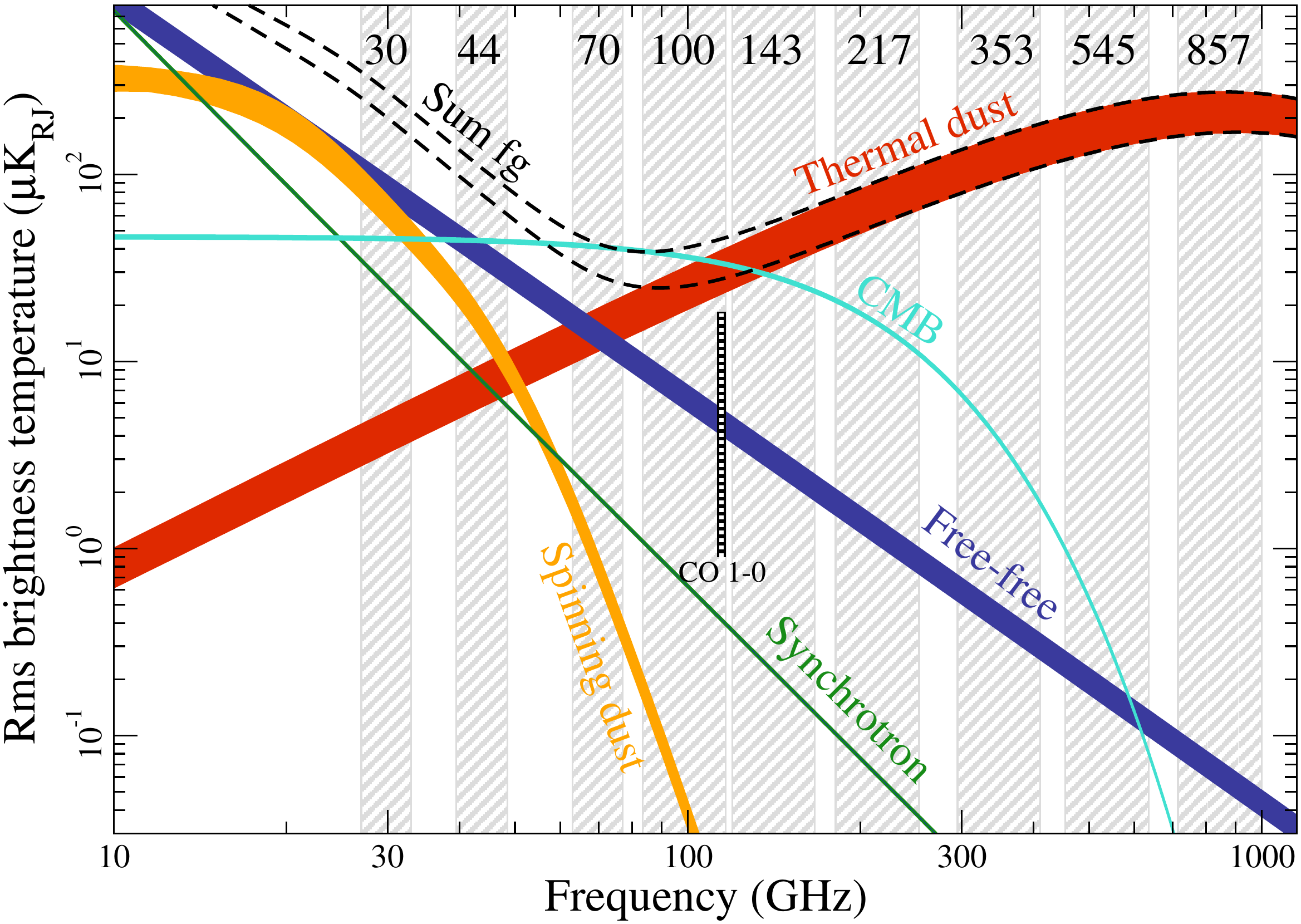}
  \caption{Illustration of the Spectral Energy Density response of each component
    in the microwave emission. The shaded bands indicate the 9 different
    observation frequencies of the Planck space observatory.
    Our goal is to create a map of each component in the sky;
    ``CMB'' emission, ``Thermal dust'' emission, and so on,
    after making assumptions about the spectral behaviour
    of each component such in this figure. We model the spectral behaviour as
    slightly different in each pixel; hence it is indicated using bands rather
    than lines. See also figure \ref{fig:spectral-index-map}.
    This figure is reproduced directly from \cite{planck2015:foregrounds}.
  }
 \label{fig:sed}
\end{figure}

\begin{figure}
  \centering
  \includegraphics[width=0.8\linewidth]{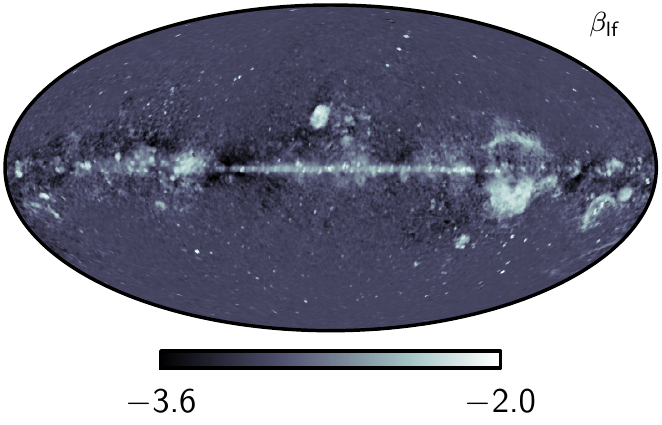}
  \caption{Example \emph{spectral index map} $\beta(\hat{n})$.
    Each pixel value corresponds to the slope in figure \ref{fig:sed}
    in that pixel for a single combined low-frequency emission component
    (using a single component to represent the ``Free-free'', ``Synchrotron''
    and ``Spinning dust'' emission types indicated in figure \ref{fig:sed}).
    In reality, each pixel is the sum of multiple slightly different spectral behaviours
    from different clouds of particles behind one another, but we instead
    work with a single compromise value for the dominating emission in the given direction.
    The \emph{mixing maps} of this paper, $q_{\nu,k}(\hat{n}_i)$, are taken
    to be proportional to $\nu^{\beta(\hat{n}_i)}$; where each component $k$ has a
    different $\beta$-map. This figure is reproduced
    directly from \cite{planck2015:foregrounds}.}
 \label{fig:spectral-index-map}
\end{figure}

\cite{joint-bayesian-compsep} describe a Bayesian model for CMB component
separation under the assumption that all observed sky maps have the
same instrumental beam and pixel resolution.  For full resolution
analysis of Planck data this is an unrealistic requirement, as the
Full-Width Half-Max (FWHM) of the beam (Point Spread Function)
span a large range, from 4.4 to 32 arc-minutes, and so one
loses much information by downgrading data to a common
resolution. In this paper we generalize the model to handle sky maps
observed with different beams and at different resolutions.

We will restrict our attention to the CMB component and diffuse
foregrounds. \cite{joint-bayesian-compsep} additionally incorporate template components
in the model for linear component separation. These are particularly useful for
dealing with point sources, where beam asymmetry is much more noted
than for the diffuse foregrounds. Recent versions of Commander
sample template amplitudes as an additional Gibbs step, rather than as
part of the linear system for component separation, so as to more easily
include a positivity constraint on such amplitudes. We will therefore ignore
templates in this paper.

The microwave sky is observed as $\Nobs$ different sky maps
$\d_\nu$ with different instrumental characteristics, and we wish to separate these
into $\Ncomp$ distinct diffuse foreground
components $\s_k$. The key to achieving this is to specify the spectral energy density (SED)
of each component; see figures \ref{fig:sed} and \ref{fig:spectral-index-map}.
\cite{joint-bayesian-compsep} describes the Gibbs sampling steps employed to
fit the SED to data. For the purposes of this paper
we will simply assume that the SED information is provided as input, in the form of
\emph{mixing maps}. Our basic assumption, ignoring any instrumental effects,
is that the true microwave emission in a direction $\hat{n}$ on the sky,
integrated over a given frequency bandpass labelled $\nu$, is given by
\begin{equation}
  \label{eq:signal-model}
f_{\nu}(\hat{n}) = \sum_k q_{\nu,k}(\hat{n}) s_k(\hat{n}),
\end{equation}
where $s_k(\hat{n})$ represents the underlying component amplitude,
and $q_{\nu,k}(\hat{n})$ represents an assumed mixing map.
We will model the CMB component simply as another diffuse component,
but note that the mixing maps are in that case spatially constant.

We do not observe $f_{\nu}(\hat{n})$ directly, but rather take as input a
pixelized sky map $\d_\nu$,
where $f_\nu$ has been convolved with an instrumental beam and then further
contaminated by instrumental noise.
To simplify notation we will employ a notation with stacked vectors, and write
\[
  \d \equiv \mat{ \d_\text{30GHz} \\ \d_\text{70GHz} \\ \vdots },
  \quad
  \s \equiv \mat{ \s_\text{cmb} \\ \s_\text{dust} \\ \vdots }.
\]
Our data observation model can then be written
\begin{equation}
  \label{eq:block-data-model}
  \d = \P \s + \n,
\end{equation}
where $\P$ is an $\Nobs \times \Ncomp$ block-matrix where each block $(\nu, k)$ projects
component $\s_k$ to the observed sky map $\d_\nu$, and $\n$ represents instrumental noise,
and is partitioned in the same way as $\d$. The noise $\n_\nu$ is a pixelized map with the same
resolution as $\d_\nu$, and assumed to be Gaussian distributed with zero mean. For our
experiments we also assume that the noise is independent between pixels, so that
$\text{Var}(\n_\nu) \equiv \N_\nu$ is a diagonal matrix, although this is not a fundamental
requirement of the method. We do however require that the matrix $\N_\nu^{-1}$
can somehow be efficiently applied to a vector.

Each component $\s_k$ represents the underlying, unconvolved field. In our implementation we work
with $\s_k$ being defined by the spherical harmonic expansion of $s_k(\hat{n})$,
truncated at some band-limit $L_k$, i.e., we assume $s_{\ell,m} = 0$ for $\ell > L_k$.
The choice of $L_k$ is essentially a part of the statistical
model, and typically chosen to match a resolution that the observed sky maps
will support. Additionally, each component $\s_k$ may have an associated Gaussian prior
$p(\s)$, specified through its covariance matrix $\S$. The role of the prior is to
introduce an assumption on the \emph{smoothness} of $\s$. The prior typically does not
come into play where the signal for a component is strong, but in regions that lacks
the component it serves to stabilize the solution, such that less noise is interpreted
as signal. Computationally it is easier to assume the same smoothness prior
everywhere on the sky, in which case the covariance matrix $\text{Var}(\s) = \S$ is diagonal
in spherical harmonic domain with elements given by the
\emph{spherical harmonic power spectrum}, $C_{k,\ell}$. However, this is not a necessary assumption, and we comment on a different type of prior in Sect.~\ref{sec:conclusions}.

The CMB power spectrum prior, given by $C_{\text{cmb},\ell}$, is particularly crucial. For the purposes of full sky
component separation one would typically not specify any prior so as to not bias the CMB. For the
purposes of estimating foregrounds, however, or filling in a CMB realization within a mask, one may
insert a fiducial power spectrum predicted by some cosmological model.

We now return to the projection operator $\P$, which projects each component to the sky.
This may be written in the form of a block matrix with each column representing a component $k$ and each
row representing a sky observation $\nu$, e.g.,
\begin{equation}
    \ve{P} = \mat{
        \ve{P}_{\text{30GHz,cmb}} & \ve{P}_{\text{30GHz,dust}} & \dots \\
        \ve{P}_{\text{70GHz,cmb}} & \ve{P}_{\text{70GHz,dust}} & \dots \\
        \vdots & \vdots & \ddots
    },
\end{equation}
with each block taking the form
\begin{equation}
  \label{eq:P}
  \ve{P}_{\nu,k} = \Y_{\nu} \B_{\nu} \widetilde{\Q}_{\nu,k}.
\end{equation}

The operator $\Y_\nu$ denotes spherical harmonic synthesis to the pixel grid employed
by $\d_\nu$. We assume in this paper an azimuthally symmetric instrumental beam for each
sky map, in which case the beam convolution operator $\B_\nu$ is diagonal in spherical harmonic
domain with elements $b_{\nu,\ell}$. This transfer function decays to zero as $\ell$ grows at a rate
that fits the band-limit of the grid.
Finally $\widetilde{\Q}_{\nu,k}$ is an operator that
denotes point-wise multiplication of the input with the mixing map $q_{\nu,k}(\hat{n})$;
computationally this should be done in pixel domain, so that
\begin{equation}
  \label{eq:Q}
  \widetilde{\Q}_{\nu,k} = \Y_{\nu,k}^T \W_{\nu,k} \Q_{\nu,k} \Y_{\nu,k},
\end{equation}
where $\Q_{\nu,k}$ contains $q_{\nu,k}(\hat{n})$ on its diagonal.  The
subscripts on the SHT operators indicate that these are defined on a
grid specific to this mixing operation.  Note that
$\widetilde{\Q}_{\nu,k}$ will cause the creation of new small-scale
modes, implying that, technically speaking, the band-limit of
$\widetilde{\Q}_{\nu,k} \s_k$ is $2 L_k$ rather than $L_k$ for a
full-resolution mixing map $q_{\nu,k}$.  For typical practical system
solving, however, the mixing matrices are usually smoother than the
corresponding amplitude maps (due to lower effective signal-to-noise
ratios), and the model may incorporate an approximation in that
$\widetilde{\Q}_{\nu,k}$ truncates the operator output at some lower
band-limit. At any rate, the grid used for $\Q_{\nu,k}$ should
accurately represent $q_{\nu,k}$ up to this band-limit. For this
purpose it is numerically more accurate to use a Gauss-Legendre grid,
rather than the HEALPix\footnote{http://healpix.jpl.nasa.gov} grid
\citep{healpix} that is usually used for $\d_\nu$. The solver of the
present paper simply treats this as a modelling detail, and any
reasonable implementation works fine as long as
$\widetilde{\Q}_{\nu,k}$ is not singular. However, to achieve
reasonable efficiency, we do require $q_{\nu,k}(\hat{n})$ to be
relatively flat, such that approximating $\widetilde{\Q}_{\nu,k}$ with
a simple scalar $q_{\nu,k}$ is a meaningful zero-order
representation. In practice we typically find ratios between the
maximum and minimum values of $q_{\nu,k}(\hat{n})$ of 1.5--3. In
general, the higher the contrast, the slower the convergence of the
solver is.

\subsection{The constrained realization linear system}
We have specified a model for the components $\s$ where the likelihood $p(\d)$ and
component priors $p(\s)$ are Gaussian, and so the Bayesian posterior is also Gaussian \citep{gibbs-jewell,gibbs-wandelt,gibbs-eriksen,joint-bayesian-compsep}, and given by
\begin{equation}
  p(\s | \d, \S) \propto e^{-\frac{1}{2} \s^T (\S^{-1} + \P^T \N^{-1} \P)^{-1} \s}.
\end{equation}
To explore this density we are typically interested in either i)
the mean vector, or ii) drawing samples from the density.
Both can be computed by solving a linear system $\A \x = \b$ with
\begin{equation}
  \label{eq:A}
  \A \equiv \S^{-1} + \P^T \N^{-1} \P
\end{equation}
and
\begin{equation}
 \b \equiv \P^T \N^{-1} \d + \P^T \N^{-1/2} \omega_1 + \S^{-1} \omega_2.
\end{equation}
The vectors $\omega_1$ and $\omega_2$ should either be
\begin{enumerate}
  \item[i)] zero, in which case the solution $\x$ will be the mean $\text{E}(\s | \d, \S)$,
    also known as the \emph{Wiener-filtered map}.
  \item[ii)] vectors of variates from the standard Gaussian
    distribution, in which case the solution $\x$ will be samples drawn $p(\s | \d, \S)$.
    We refer to such samples as the \emph{constrained realizations}.
\end{enumerate}
For convenience we refer to the system as the constrained realization (CR) system
in both cases. The computation of the right-hand side $\b$ is straightforward
and not discussed further in this paper; our concern is the efficient solution of the
linear system $\A \x = \b$.

For current and future CMB experiments, the matrix $\A$ is far too
large for the application of dense linear algebra. We are however able
to efficiently apply the matrix to a vector, by applying the operators
one after the other, and so we can use iterative solvers for linear
systems.  Since $\A$ is symmetric and positive definite, the
recommended iterative solver is the {\em Conjugate Gradients} (CG)
method \citep[see][for a tutorial]{CGWithoutAgonizingPain}.  One starts
out with some arbitrary guess, say, $\x_1 = \ve{0}$. Then,
for each iteration, a {\em residual} $\r_i = \b - \A \x_i$ is computed,
and then this residual is used to produce an updated iterate $\x_{i+1}$ that
lies closer to the true solution $\x_\text{true}$.

Note that the residual $\r_i$, which is readily available, is used as a proxy
for the {\em error}, $\e_i = \x_\text{true} - \x_i$, which is unavailable
as we do not know $\x_\text{true}$. The key is that since $\A \x_\text{true} = \b$,
\[
\r_i = \b - \A \x_i = \A (\x_\text{true} - \x_i) = \A \e_i,
\]
and since $\A$ is linear, reducing the magnitude of $\r_i$ will also lead to
a reduction in the error $\e_i$.
In a production run the error $\e_i$ is naturally unavailable, but
during development and debugging it is highly recommended to track it.
This can be done by generating some $\x_\text{true}$, then generate
$\b = \A \x_\text{true}$, and track the error $\e_i$ while running the
solver on this input. The benchmark results presented in this paper
are generated in this manner.

The number of iterations required by CG depends on how uniform (or
clustered) the eigenspectrum of the matrix is. Therefore the main
ingredient in a linear solver is a good {\em preconditioner} which
improves the eigenspectrum.  A good preconditioner is a symmetric,
positive definite matrix $\M$ such that $\ve{M} \x$ can be quickly
computed, and where the eigenspectrum of $\ve{M} \A$ is as flat as
possible, or at least clustered around a few values.  Intuitively,
$\M$ should in some sense approximate $\A^{-1}$.

\section{A full sky preconditioner based on pseudo-inverses}
\label{sec:pseudo-inverse-preconditioner}
\subsection{On pseudo-inverses}
The inspiration to the solver presented below derives from the literature on the numerical solution of
\emph{Saddle-Point systems}.
Our system $\A$, in the form given in  in equation \eqref{eq:A}, is an instance of a
so-called \emph{Schur complement}, and approximating the inverse of such Schur complements
plays a part in most solvers for Saddle-Point systems.
An excellent review on the solution of saddle-point systems is given
in \cite{saddle-point-review}. The technique we will use was
originally employed by \cite{elman:1999} and \cite{elman:2006}, who
use appropriately scaled \emph{pseudo-inverses} to develop
solvers for Navier-Stokes partial differential equations.

The \emph{pseudo-inverse} is a generalization of matrix inverses to rectangular
and/or singular matrices.  In our case, we will deal with an
$m \times n$-matrix $\U$ with linearly independent columns and $m>n$, in which
case we have that the pseudo-inverse is given by
\begin{equation*}
  \label{eq:pseudo-inv-fullrank}
  \U^+ \equiv (\U^T\U)^{-1}\U^T.
\end{equation*}
Note that in this case, $\U^+$ is simply the matrix that finds the least-squares solution to the
linear system $\U \x = \b$. This matrix has the property that
\[
\U^+ \U = (\U^T\U)^{-1}\U^T \U = \I,
\]
and so it is a \emph{left inverse}. Unlike the real inverse however, $\U \U^+ \ne \I$.
This follows from $\U^+$ having the same shape as $\U^T$,
so that $\U \U^+$ is necessarily a singular matrix.
For small matrices, $\U^+$ can be computed by using the QR-decomposition.
Note that in the case that $\U$ does \emph{not} have full rank, the pseudo-inverse
has a different definition and should be computed using the Singular Value Decomposition (SVD);
the definition above is however sufficient for our purposes.

\subsection{Full-sky, single component, without prior}
\label{sec:full-sky}

For the first building block of our preconditioner we start out with a much simpler
linear problem. Assuming a model with a single sky map $\d$,
a single component, no mixing maps, and no prior,
the linear system of equation \eqref{eq:A} simply becomes
\begin{equation}
\B \Y^T \N^{-1} \Y \B \x = \b.
\end{equation}
Note that $\Y^T$ is \emph{not} the same as
spherical harmonic analysis, which in our notation reads $\Y^T \W$.
Without the weight matrix $\W$, $\Y^T$ represents
\emph{adjoint spherical harmonic synthesis}, and simply falls out algebraically
from the transpose sky-projection $\P^T$.

\begin{figure}
  \centering
  \includegraphics[width=1\linewidth]{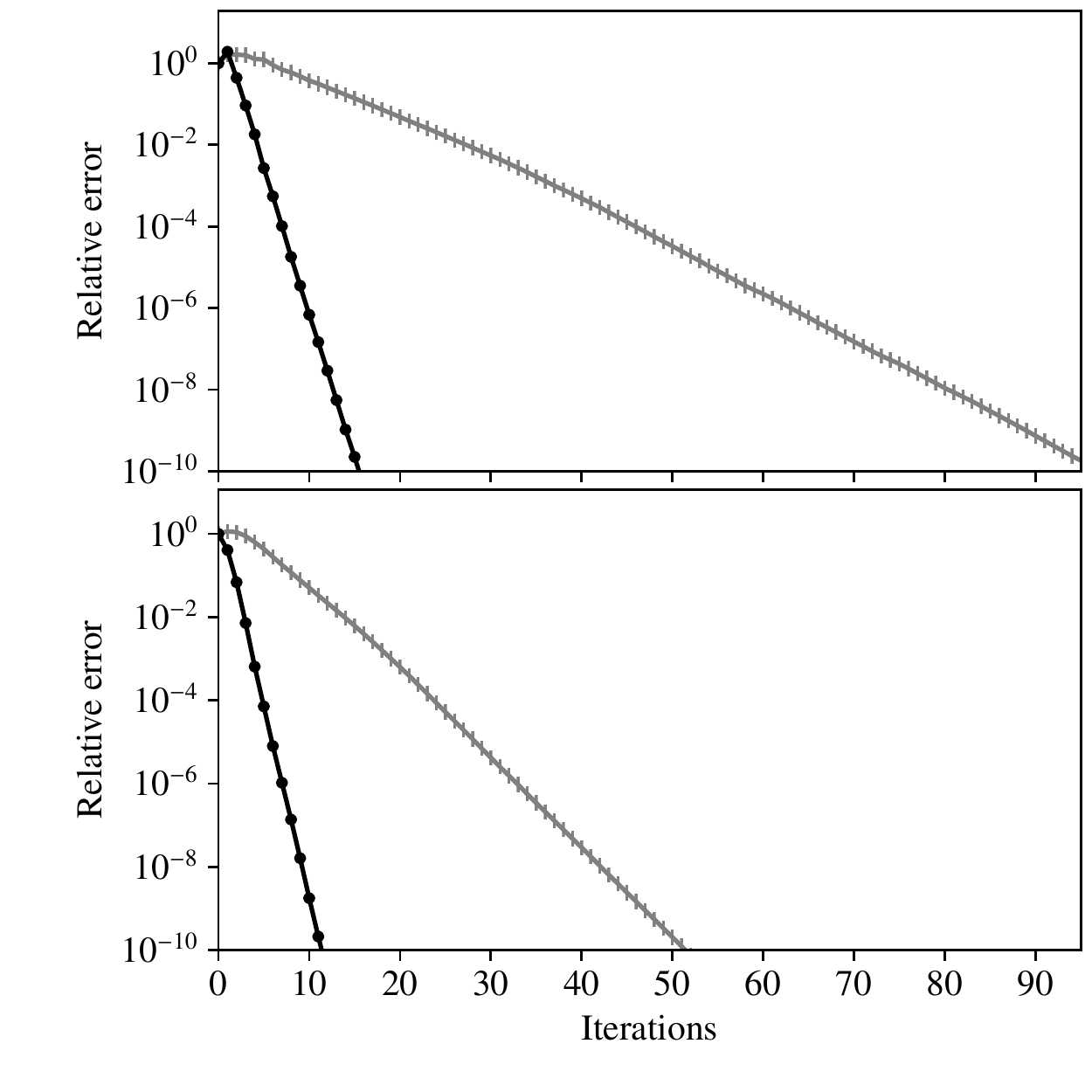}
  \caption{Convergence of preconditioner of Eq. \eqref{eq:simplest-preconditioner}
    (black circles) compared to a simple diagonal preconditioner (gray ticks).
    In this case, we fit a single CMB component to a single 143 GHz Planck band without specifying a prior.
    {\it Top panel}: Convergence rate for an unmodified RMS map with $\sigma_\text{max}/\sigma_\text{min} = 24$.
    {\it Bottom panel}: Convergence rate for a modified RMS map where we have added synthetic noise to the
    1\% pixels with least noise, in order to bring $\sigma_\text{max}/\sigma_\text{min}$ down to 7.5.
    With a higher contrast in the RMS map the pseudo-inverse preconditioner brings more of
    an improvement relative to the diagonal preconditioner.
    In the  case of a flat RMS map the two preconditioners are numerically
    equivalent. See Sect.~\ref{sec:benchmarks} for details on the benchmark setup
    and the diagonal preconditioner.
  }
 \label{fig:benchmark_single}
\end{figure}

Since the basis of $\x$ is in spherical harmonic domain, the beam convolution operator $\B$ is diagonal
and trivially inverted. The remaining matrix $\Y^T \N^{-1} \Y$ is diagonal in pixel domain, so
if only $\Y$ had been invertible we could have solved the system directly.
Inspired by \cite{elman:1999}, we simply pretend that $\Y$ is square, and
let $\Y^T \W$ play the role of the inverse, so that
\begin{equation}
\label{eq:simplest-preconditioner}
(\B \Y^T \N^{-1} \Y \B)^{-1} \approx \B^{-1} \Y^T \W \N \W \Y \B^{-1}.
\end{equation}
We stress that because $\Y$ is not \emph{exactly} invertible and $\Y^T \W$ is a pseudo-inverse,
this is only an approximation which should be used as a preconditioner inside
an iterative solver. For the HEALPix grid in particular,
$\Y^T \W$ is rather inaccurate and we only have $\Y^T \W \Y \approx \I$;
it is however a very good approximation, as can be seen in Fig. \ref{fig:benchmark_single}.

\subsection{Full-sky, multiple components, with priors}
\label{sec:full-sky-multi-comp}

\begin{figure}
  \centering
  \includegraphics[width=\linewidth]{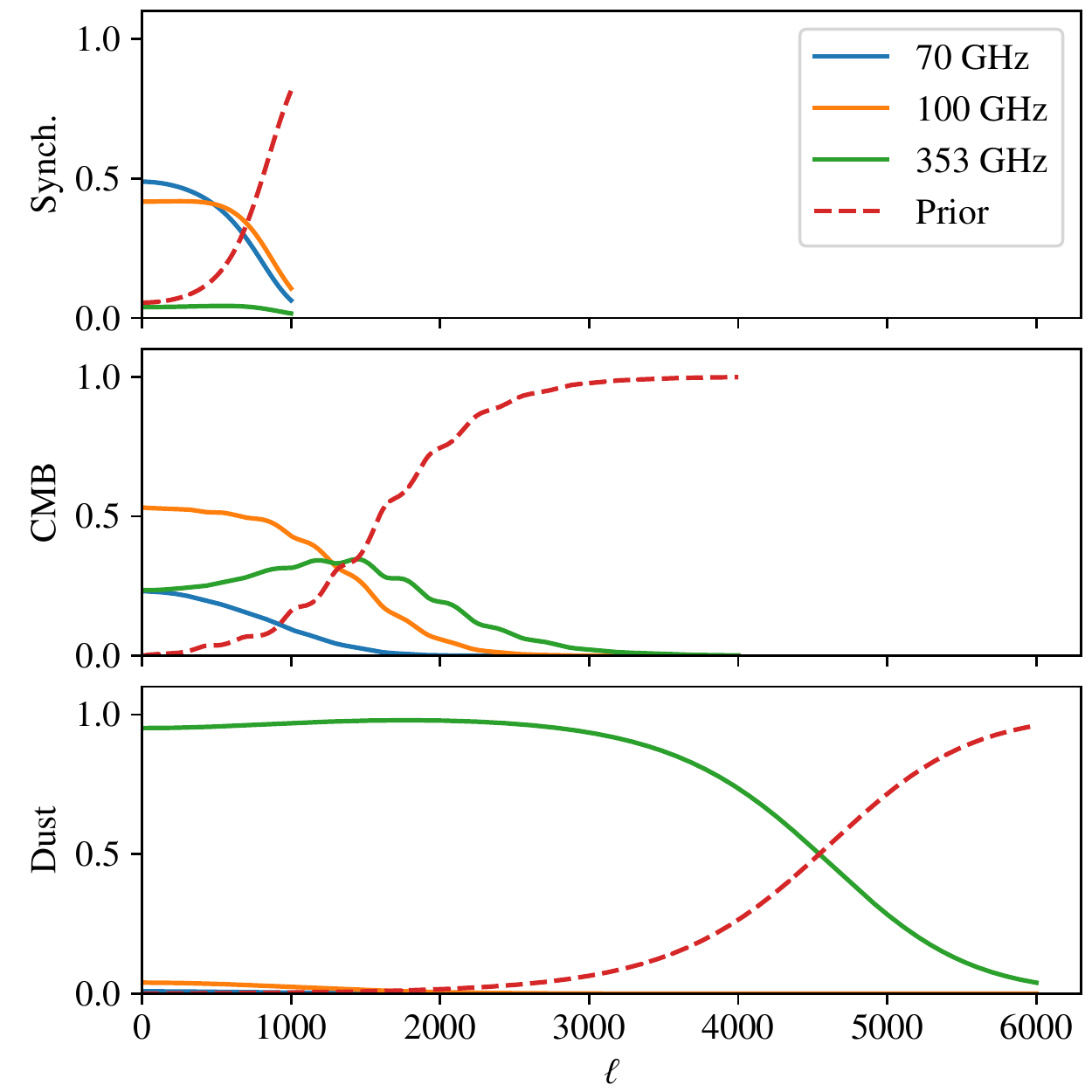}
  \caption{Visualization of the matrix $\U$ in an example setup. For each component $k$
    we plot the coefficients along the corresponding column of $\U$, normalized so that
    the sum is $1$ for each $\ell$. Note how the  CMB has most support
    from the 100 GHz band for low $\ell$, then gradually switches to the
    353 GHz band and finally the prior as $\ell$ increases.
    }
  \label{fig:U}
\end{figure}

Next, we want to repeat the above trick for a case with multiple sky maps $\d_\nu$,
multiple components $\x_k$, and a prior. We start by assuming that the mixing operators $\WQ_{\nu,k}$ can be reasonably approximated
by a constant,
\[
\WQ_{\nu,k} \approx q_{\nu,k} \I,
\]
although, as noted above, this is only a matter of computational efficiency of the preconditioner.
The optimal value for $q_{\nu,k}$ is given in Appendix
\ref{appendix:approximating-mixing-map}.  In the following derivation
we will simply assume equality in this statement, keeping in mind that
all the manipulations only apply to the preconditioner part of the CG
search, and so does not affect the computed solution.

\definecolor{emphcolor}{rgb}{0.8,0.0,0.0}
\begin{figure*}
  \begin{minipage}{0.6\textwidth}
  \centering
  \includegraphics[width=\linewidth]{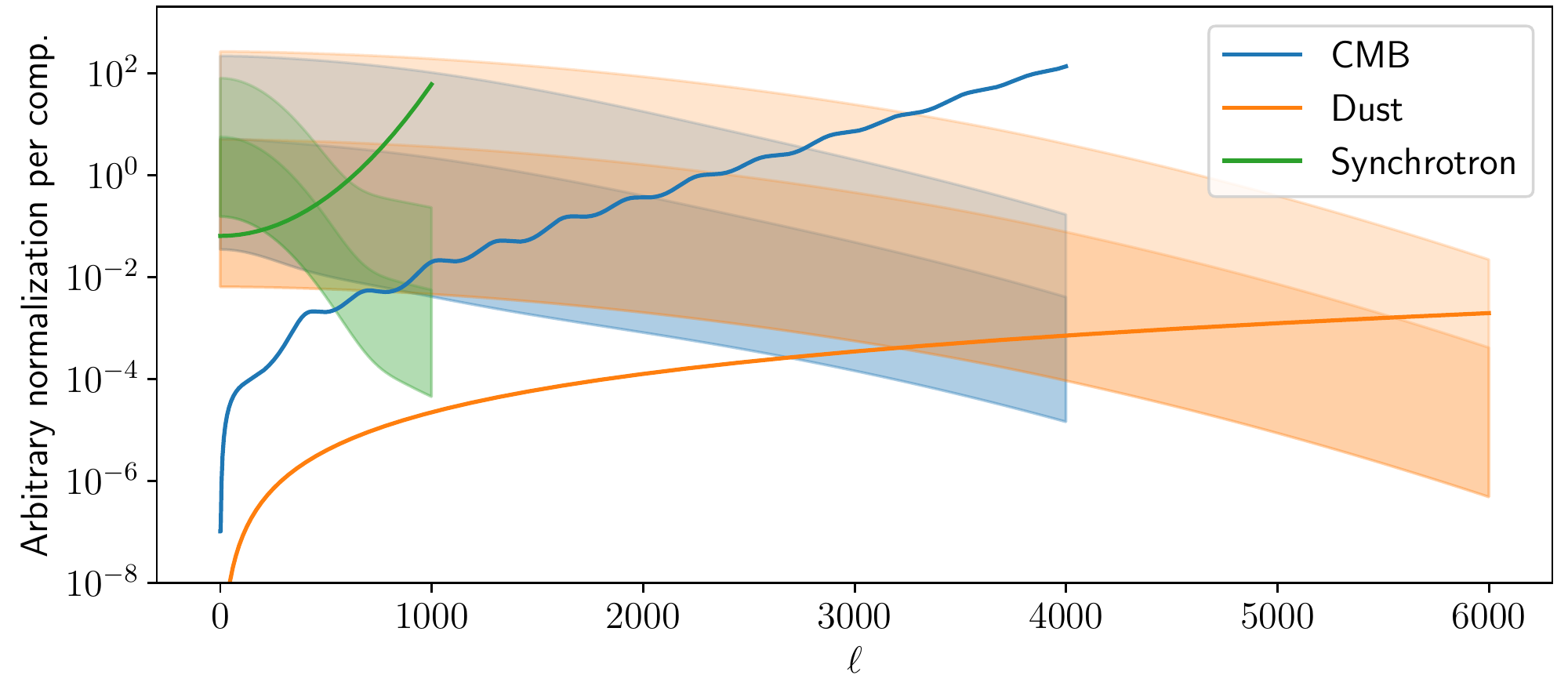}
  \end{minipage}
  \begin{minipage}{0.3\textwidth}
    \centering
    \normalsize
     {
       \setlength{\extrarowheight}{2mm}
    \begin{tabular}{l|cc}
      & Inside mask & Outside mask \\\hline
      Low $\ell$ & {$\color{emphcolor} \A \approx \S^{-1}$} & $\A \approx \P^T \N^{-1} \P$  \\
      High $\ell$ & $\A \approx \S^{-1}$ & $\A \approx \S^{-1}$
    \end{tabular}
    }
  \end{minipage}
  \caption{The significance of the prior term versus the inverse-noise term of $\A$.
    {\bf Left pane:} The significance as a function of scale. The coefficients of
    the prior term $\S_k^{-1}$ (solid lines) and the
    inverse-noise term $\sum_\nu \P^T_{\nu,k} \N_\nu^{-1} \P_{\nu,k}$ (bands),
    whose sum make up a diagonal block $\A_{k,k}$ of $\A$, in a simple example setup.
    For small $\ell$ (large scales) the inverse-noise term dominates, while
    for larger $\ell$ (small scales) the prior term dominates.
    The bands extend between the minimum and maximum matrix coefficient values,
    and indicates the amount of spatial variation that is present and which
    cannot be represented well in spherical harmonic domain.
    For the lighter bands we have assumed all 9 sky maps from the Planck space observatory,
    while for the darker bands we have added small amounts of synthetic noise to
    the $1\%$ least noisy pixels, effectively thresholding the RMS map in a way
    that does not meaningfully impact the solution of the system while having a large
    effect on the matrix. Displayed is the full sky case; in the presence of a mask,
    the bands would simply have extended to zero as the inverse-noise term
    would become singular.
    {\bf Right pane:} A schematic setup of which term dominates $\A$ in different regimes.
    The pseudo-inverse preconditioner of Sect. \ref{sec:pseudo-inverse-preconditioner}
    automatically resolves the low-$\ell$ vs. high-$\ell$ split, taking into account
    cross-component couplings. To handle the crucial low-$\ell$ regime inside the mask (red) the extension
    described in Sect. \ref{sec:mask} is required.}
  \label{fig:regimes}
\end{figure*}

First note that the matrix of equation
\eqref{eq:A} can be written
\[
\A = \mat{ \P^T & \I } \mat{ \N^{-1} & \\ & \S^{-1} } \mat{ \P \\ \I }.
\]
Writing the system in this way makes it evident that in the Bayesian framework,
the priors are treated just like another ``observation'' of the components.
These ``prior observations'' play a bigger role for parts of the solution where
the instrumental noise is high, and a smaller role where the instrumental noise is low
(see Fig. \ref{fig:U}).

The idea is now to further rearrange and rescale the system in such a way that
all information that is not spatially variant is expressed in the projection
matrix on the side, leaving a unit-free matrix containing only spatial
variations in the middle. Since $\S^{-1}$ is assumed to be diagonal, it is trivial to
factor it and leave only the identity in the center matrix. Turning to
the inverse noise term $\N^{-1}$, we find experimentally that rescaling
with a scalar performed well. Specifically, we define
\[
\widetilde{\N}_\nu^{-1} \equiv \alpha_\nu^{-2} \Y_\nu^T \N^{-1}_\nu\Y_\nu,
\]
where $\alpha$ takes the value that minimizes $\| \widetilde{\N}_\nu^{-1}(\alpha_\nu) - \I \|_2$.
Computing $\alpha_\nu$ is cheap, and its optimal value is derived in
Appendix \ref{appendix:approximating-inverse-noise-map}.
A similar idea is used by \cite{elsner-messaging},
as they split $\N^{-1}$ into a spherical harmonic term and a
pixel term, but their split is additive rather than
multiplicative. Our system can now be written
\begin{equation}
  \label{eq:A-for-pseudoinv}
  \A = \U^T \T \U,
\end{equation}
where $\T$ contains the unit-free spatial structure of the system, and
$\U$ the spatially invariant, but $\ell$-dependent, structure of the system:
\[
  \T \equiv  \mat{\widetilde{\N}^{-1} &  \\   & \I },
  \quad
  \U \equiv
  \mat{
    \alpha \B \WQ \\ \S^{-1/2}
  }.
\]
In order to elucidate the block structure of these matrices we write out an example
with 3 bands and 2 components:
\[
\small
 \T =  \mat{
    \widetilde{\N}^{-1}_1  &          &          & & \\
              & \widetilde{\N}^{-1}_2  &          & & \\
              &          & \widetilde{\N}^{-1}_3 & & \\
              &          &          & & \\
              &          &          & \I  & \\
              &          &          & & \I  \\
 },
 \quad
  \U = \mat{
    \alpha_1 \B_1 \WQ_{1,1} & \alpha_1 \B_1 \WQ_{1,2} \\
    \alpha_2 \B_2 \WQ_{2,1} & \alpha_2 \B_2 \WQ_{2,2} \\
    \alpha_3\B_3 \WQ_{3,1} & \alpha_3 \B_3 \WQ_{3,2} \\
    \S_1^{-1/2}   &      \\
          &  \S_2^{-1/2}
  }.
\]
Under the assumption that $\WQ_{\nu,k} = q_{\nu,k} \I$, the matrix
$\U$ is block-diagonal with small blocks of variable size when seen in
$\ell$- and $m$-major ordering. Specifically, the entries in the
``data-blocks'' $(\nu, k')$ in the top part of $\U$ are given by
\begin{equation}
  \label{eq:U-data}
U_{(\ell,m,\nu),(\ell',m',k')} = \alpha_\nu b_{\nu,\ell} q_{\nu,k'} \delta_{\ell,\ell'} \delta_{m,m'},
\end{equation}
and the entries in the ``prior-blocks'' $(k, k')$ in the bottom part of $\U$
are given by
\begin{equation}
  \label{eq:U-prior}
U_{(\ell,m,k),(\ell',m',k')} = C_{k,\ell}^{-1/2} \delta_{k,k'} \delta_{\ell,\ell'} \delta_{m,m'}.
\end{equation}
In the event that no prior is present for a component, the corresponding block can
simply be removed from the system, or one may equivalently set $C_{k,\ell}^{-1/2} = 0$.

At this point we must consider the multi-resolution nature of our setup.
Our model assumes a band-limit $L_k$ for each component, so that
each component has $(L_k + 1)^2$ corresponding columns in $\U$, and, if a
prior is used, an additional corresponding $(L_k + 1)^2$ rows.
Each sky observation has a natural band-limit $L_\nu$
where the beam transfer function $b_{\nu,\ell}$ has decayed so much that
including further modes is numerically irrelevant, and so each band
has $(L_\nu + 1)^2$ corresponding rows in $\U$.
When we view $\U$ in the block-diagonal $\ell$- and $m$-major ordering,
the block sizes are thus variable: Each band-row only participates
for $\ell \le L_\nu$, and each component-column and component-row only
participates for $\ell \le L_k$.
A way to see this is to consider that the first blocks for $\ell = 0$
have size $(\Nobs + \Ncomp)\times\Ncomp$ for all $m$; then, as $\ell$ is increased past some
$L_\nu$ or $L_k$, corresponding rows or columns disappears from the blocks.

In code it is easier to introduce appropriate zero-padding for $\ell > L_k$ and $\ell > L_\nu$
so that one can work with a single constant block size. Either way,
the numerical results are equivalent. Since $\U$ consists of many small blocks
on the diagonal computing the pseudo-inverse is quick, and the memory use
and compute time for constructing $\U^{+}$ scales as $O(\Ncomp^2 \Nobs L^2)$.

With this efficient representation of $\U^{+}$ in our toolbox, we again use the trick of \cite{elman:1999}
to construct a preconditioner on the form
\begin{equation}
  \label{eq:fullmodel-preconditioner}
\A^{-1} = (\U^T \T \U)^{-1} \approx \U^{+} \T^{-1} (\U^{+})^T.
\end{equation}
The lower-right identity blocks of $\T$ requires no inversion.
The $\widetilde{\N}^{-1}_\nu$-blocks in the upper-left section of $\T$
can be approximately inverted using the technique presented in Sect. \ref{sec:full-sky},
so that
\[
(\widetilde{\N}^{-1}_\nu)^{-1} \approx
\alpha_\nu^{2} \Y_\nu^T \W_\nu \N_\nu \W_\nu \Y_\nu.
\]
Let $\T^+$ denote $\T^{-1}$ approximated in this manner, and the resulting preconditioner is
\begin{equation}
  \M_\text{PI} = \U^{+} \T^+ (\U^{+})^T.
\end{equation}
Computationally speaking, this has the optimal operational form: The pseudo-inverse $\U^+$ and its transpose
can be applied simply by a series of small matrix-vector products in parallel using the pre-computed
blocks, while application of $\T^+$ as an operator requires $2 \Nobs$ SHTs.

\begin{figure}
  \centering
  \includegraphics[width=1\linewidth]{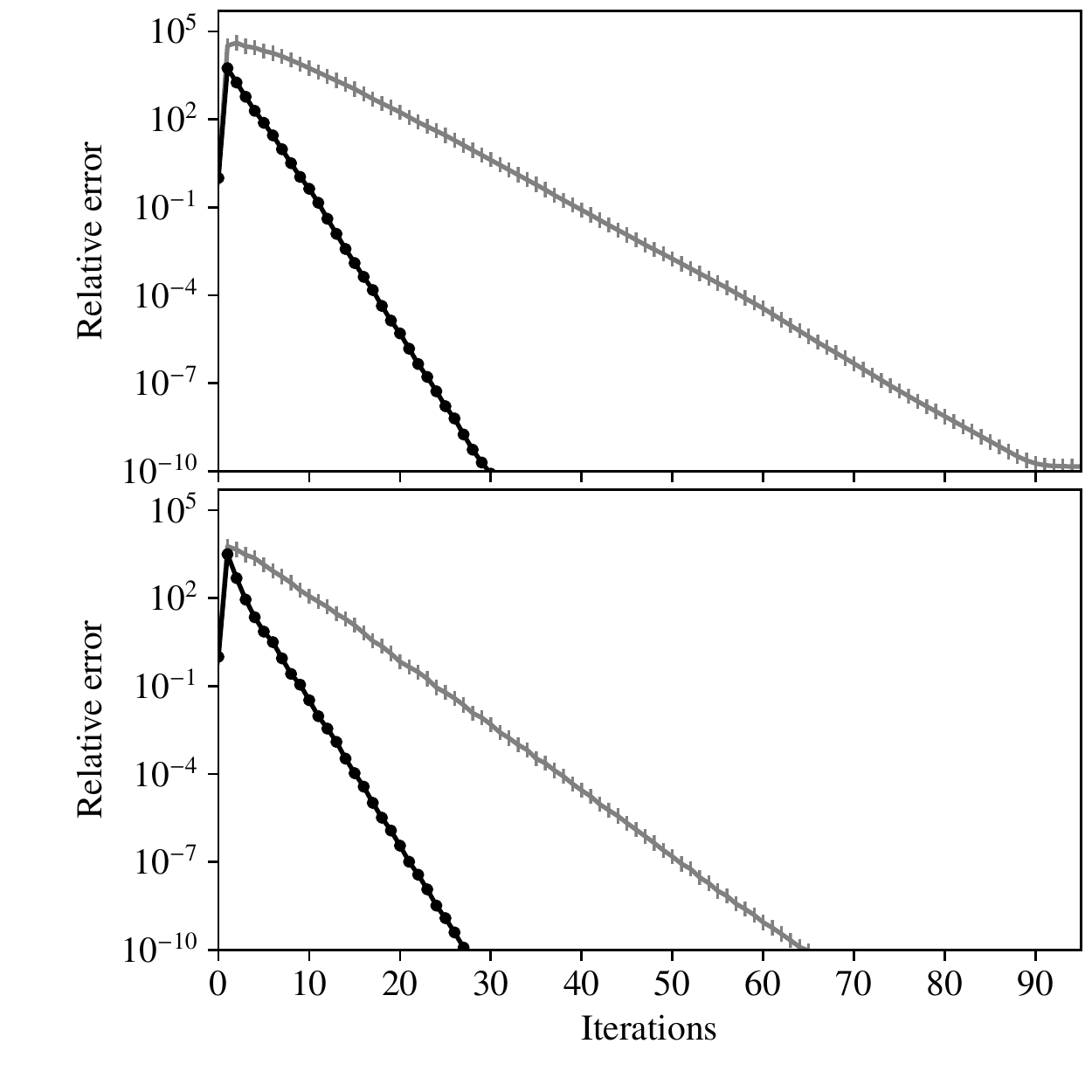}
  \caption{Convergence of the pseudo-inverse preconditioner for a
    full-sky component separation model. The pseudo-inverse
    preconditioner (black circles) is shown together with a block diagonal
    preconditioner (gray ticks). We fit the model against data from all nine
    bands of Planck, each with a different instrumental beam.
    The model has three microwave components:
    i) A synchrotron component, band-limit $L_\text{synch}=1000$, a Gaussian
    prior $C_{\text{synch},\ell}$ with FWHM of 30 arc-minutes and with
    signal-to-noise ratio of unity at $\ell=350$.
    ii) A CMB component, band-limit $L_\text{cmb}=4000$, a $\Lambda$CDM
    power spectrum prior, with signal-to-noise ratio of unity around $\ell=1600$.
    iii) A dust component, band-limit $L_\text{dust}=6000$, with signal-to-noise
    ratio of unity around $\ell=4500$.
    These parameters all refer to the full-resolution model, which was then degraded
    to lower resolution for these runs as detailed in Sect. \ref{sec:benchmarks}.
    {\it Top panel:} Using variable mixing maps for the dust component, where
    the ratio of maximum and minimum  values ranges from 1.5 to 3. The mixing
    maps for the synchrotron components and the CMB components are flat.
    {\it Bottom panel:} All mixing maps are flat.
  }
 \label{fig:benchmark_compsep}
\end{figure}

So far we have only considered the pseudo-inverse as an algebraic trick, but some
analytic insights are also available. First we look at two special cases.
The preconditioned system can be written
\begin{align*}
  \M_\text{PI} \A 
  = (\U^T\U)^{-1}\U^T \T^+ \U (\U^T\U)^{-1} \U^T \T \U.
\end{align*}
First, note that if the real model indeed has a flat inverse-noise
variance map, then $\T^+ = \T = \I$ and the preconditioner is perfect,
$\M\A = \I$.  Second, consider the case in which the inverse-noise variance maps are
\emph{not} flat, but that rather 1) $\Ncomp = \Nobs$; 2) there are no
priors; and 3) the band-limit of each sky observation matches the
one of its dominating component. In this case, $\U$ is square and
invertible, and $(\U^T\U)^{-1} \U^T = \U^{-1}$, and thus $\M\A =
\U^{-1} \T^+ \T \U \approx \I$, using the very good approximation
developed in Sect.~\ref{sec:full-sky}.

In the generic case, $\T \ne \I$, and $\U$ has more rows than columns.
First note that for an other equivalent model with constant
inverse-noise variance maps, the corresponding system matrix is
$\A_\text{F} \equiv \U^T \U$.  Further, we can define a system which
acts just like $\A$ in terms of effects that are spatially invariant,
but which has the inverse effect when it comes to the
scale-free spatial variations: $\A_\text{SPI} \equiv \U^T \T^{+} \U$.
With these definitions we have
\begin{align*}
  \M_\text{PI} \A &= \A^{-1}_\text{F} \A_\text{SPI} \A^{-1}_\text{F} \A.
\end{align*}
The operator above applies the spatially variant effects once
forward and once inversely, and the spatially invariant effects twice forward and twice inversely.
Everything that is done is also undone, and in this sense $\M_\text{PI} \A$ can be said to approximate $\I$.

To see where the approximation breaks down, one must consider what
$\A_\text{SPI}$ actually represents, which is a linear combination of
the spatial inverses of the inverse-noise maps and the priors.  The
$\U$ matrix gives more weight to a band with more information for each
component. For instance, in a Planck-type setup that includes a
thermal dust component, the spatial inverse of the 857 GHz band will
be given strong weight, while the spatial inverse of 30 GHz will be
entirely neglected, as desired. Likewise, the prior terms will be
given little weight in the data-dominated regime at low multipoles,
and then gradually be introduced as the data becomes noise-dominated.

However, the weights between the spatial inverses ignore spatial
variations, and only account for the average within each multipole
$\ell$. Thus the preconditioner only works well in when faced with
modest spatial variations in the inverse-noise maps.  This is
illustrated in Fig.~\ref{fig:regimes}. For the spatially invariant
part of the preconditioner, the crossover between the inverse-noise
dominated regime and the prior-dominated regime must happen at a
single point in $\ell$-space, whereas in reality this point varies based on
spatial position for $\ell$ in the mid-range where the prior lines are
crossing the inverse-noise bands.

Figure~\ref{fig:benchmark_compsep} summarizes the performance of the
above preconditioner in a realistic full sky component separation
setup in terms of iteration count. The analysis setup corresponds to a
standard nine-band Planck data set in terms of instrumental noise
levels and beam characteristics \citep{planck2015:mission}. Of course,
it should be noted that the pseudo-inverse preconditioner requires
extra SHTs compared to a standard diagonal preconditioner, and so
requires some more compute time per iteration.  This penalty is highly
model-specific as it depends on $\Ncomp$, $\Nobs$ and the number of
non-constant mixing maps $\Nmix$.  For this particular model with two
flat-mixing components and one variable-mixing component, each
multiplication with $\A$ requires $2(2 \Nmix+\Nobs) = 54$ SHTs,
whereas application of the pseudo-inverse preconditioner requires
$2\Nobs = 18$ SHTs.  If all components has the same resolution this
translates to each iteration of the pseudo-inverse solver taking 33\%
longer than the block diagonal preconditioner, and with a third of the
iterations needed, we end up with a total run-time reduction of
60\%. In the model benchmarked, the synchrotron component with a flat
mixing maps has much lower resolution than the dust component with a
variable mixing map, and so the speedup is somewhat larger.

\section{Constrained realization under a mask}
\label{sec:mask}

So far we have only considered the full sky case. In many practical
applications, we additionally want to mask out parts of the sky,
either because of missing data, or because we do not trust our
statistical model in a given region of the sky. Within such scenarios,
it is useful to distinguish between two very different cases:
\begin{itemize}
\item[i)] Partial sky coverage, where only a small patch of the sky has been observed, and we wish
  to perform component separation only within this patch. The typical use-case is for
  this setup is ground-based or sub-orbital CMB experiments.

\item[ii)] Natively full-sky coverage, but too high foregrounds in a
  given part of the sky to trust our model. In this case, one often
  masks out part of the sky, but still seeks a solution to the system
  under the mask, constrained by the observed sky at the edges of the
  mask and determined by the prior inside the mask.  By ignoring data
  from this region we at least avoid that the CMB component is
  contaminated by foreground emission. Of course, the solution will
  not be the true CMB sky either, but it will have statistically
  correct properties for use inside of a Gibbs sampler
  \citep{gibbs-jewell,gibbs-wandelt,gibbs-eriksen,joint-bayesian-compsep}.
\end{itemize}

\noindent
We expect the solver developed in the previous section to work well in case i)
given appropriate modifications, but leave such modifications for future work,
and focus solely on case ii) in what follows.

\subsection{Incorporating a mask in the model while avoiding ringing effects}
Recall from Sect.~\ref{sec:model} that our data model reads
\[
\d = \Y_{\nu} \B_{\nu} \widetilde{\Q}_{\nu,k} \s + \n,
\]
where $\d$ are the pixels of the observed sky maps.  In
\cite{gibbs-eriksen} and \cite{multigrid} a mask is introduced into
the model by declaring that these pixels are missing from $\d$, or,
equivalently, that $\N^{-1}$ is zero in these pixels.  This is
straightforward from a modelling perspective, but has an inconvenient
numerical problem: Ideally, we want to specify the beam operator $\B$
using a spherical harmonic transfer function $b_\ell$, but, because
$b_\ell$ must necessarily be band-limited, $\B$ exhibits ringing in
its tails in pixel basis. Specifically, in pixel space the beam
operator first exhibits an exponential decay, as desired, but then
suddenly stops decaying before it hits zero. At this point, it starts
to observe the entire sky through the ringing ``floor'' \cite[see
  figures in][]{multigrid}, and it becomes non-local.  When the
signal-to-noise ratio of the data in question is high enough compared
to such numerical effects, the model will try to predict the signal
component $\s$ within the mask through deconvolution of the pixels at
the edge of the mask, regardless of their distance. This causes a
major complication for all solvers of this type, and in
\cite{multigrid} we had to carefully tune the solver to avoid this
ringing effect.

In the present solver we side-step this problem by introducing the
mask in the mixing maps $\Q_{\nu,k}$, rather than in the noise
model. The sky is then split cleanly into one set of pixels outside
the mask that hits the full matrix $\S^{-1} + \P^T \N^{-1} \P$, and
another set of pixels (those under the mask) that only hits the prior
term $\S^{-1}$.

The statistical interpretation of this mask model is that to pretend
that a massive screen has been installed far away in the universe,
physically shielding the microwave radiation in the region of the
mask. This model is of course not physically meaningful, but the
numerical difference is only evident in how quickly the mask takes
effect, within a region spanning one beam-width around the mask
border.  Since the masks in use are mainly constructed by rules of
thumb and by looking at residual maps, the difference between this
mask model and actually removing pixels from $\d$ is statistically irrelevant.

Another advantage of this mask model is that it enables the use of
different masks for different microwave components, although we have
not yet tried this feature of our solver on a real analysis. A typical
use-case could be joint estimation of CMB and cosmic infrared
background (CIB) fluctuations, which typically would require different
effective masks. Conversely, one disadvantage is that the same set of
masks applies to all input sky maps.

\subsection{Independently solving for signals under a mask}

\begin{figure}
  \centering
  \includegraphics[width=1\linewidth]{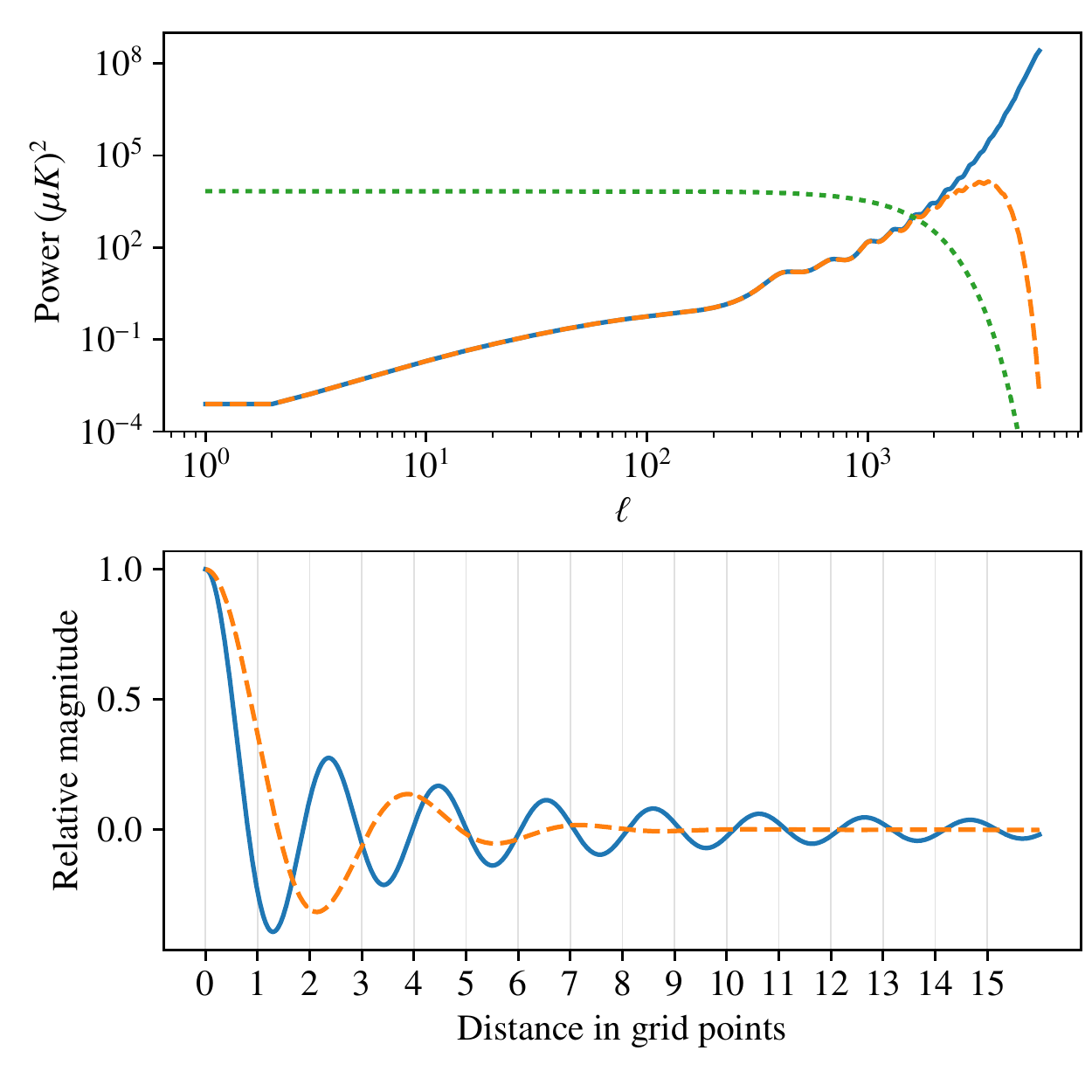}
  \caption{Harmonic filtering of the mask-restricted system $\Z \S^{-1} \Z^T$. {\it Top panel:}
    The inverse CMB power spectrum $1 / C_\ell$ (solid blue), on the diagonal of the $\S^{-1}$ matrix behaves as $\ell^2$ near the beginning,
    increasing in steepness to $\ell^8$ at $\ell=6000$. It crosses the diagonal of the inverse-noise term $\N^{-1}$
    (dotted green) at around $\ell=1600$.
    Above this point the system becomes prior-dominated and the pseudo-inverse preconditioner
    works well both inside and outside the mask. Since we do not need the dedicated mask solver to solve
    for high $\ell$, we apply a filter $r_\ell$ as described in Sect. \ref{sec:multigrid},
    resulting in a filtered prior $r_\ell^2 / C_\ell$ (dashed orange).
    {\it Bottom panel:} The $\S^{-1}$ operator in pixel domain. The plot displays
    $(\Y \S^{-1} \Y^T)_{ij}$ as a function of the distance between grid points $i$ and $j$.
    Both the unfiltered operator $1/C_\ell$
    (solid blue) and the low-pass filtered $r_\ell^2 / C_\ell$ (dashed orange) have a ``Mexican hat'' or Wavelet-like
    azimuthally symmetric shape. The low-pass filter ensures that the oscillations decay quickly,
    making the system easier to solve in pixel domain than the unfiltered version.}
 \label{fig:mask_solver}
\end{figure}

Consider the schematic description of each regime in the right pane of
Fig.~\ref{fig:regimes}. The pseudo-inverse preconditioner, which we
will denote $\M_\text{PI}$, automatically finds a good split per
component for the low- and high-$\ell$ regimes, respectively, but is,
in the same way as a block-diagonal preconditioner, blind to the
different regimes inside and outside of the mask. As a result, it
performs poorly for \emph{large scales inside the mask}, i.e., for
multipoles lower than the point of unity signal-to-noise ratio shown
in Fig.~\ref{fig:mask_solver}.

To solve this, we supplement the pseudo-inverse preconditioner with a
second preconditioner that is designed to work well only inside the
mask, where it is possible to simplify the system.  Let $\Z$ denote
spherical harmonic synthesis to the pixels within the mask only; that
is, we first apply $\Y$, and then select only the masked pixels.
Then, building on standard domain decomposition techniques, a
preconditioner that provides a solution only within the masked region
is given by
\begin{equation}
  \label{eq:M_mask}
\M_\text{mask} = \Z^T (\Z \A \Z^T)^{-1} \Z = \Z^T (\Z \S^{-1} \Z^T)^{-1} \Z.
\end{equation}
We will develop a solver for the inner \emph{mask-restricted system}
in Sect.~\ref{sec:multigrid}.
%
%
For now we assume that we we can efficiently apply a suitable approximation
of $(\Z \S^{-1} \Z^T)^{-1}$ to a vector.
It then remains to combine $\M_\text{PI}$ and $\M_\text{mask}$. The simplest possible way of
doing this is simply to add them together, $\M_\text{add} \equiv
\M_\text{PI} + \M_\text{mask}$, and we find experimentally that this
simple combination performs well for our purposes.

This may however break down for more demanding models. We found that
the most important feature in how well $\M_\text{add}$ works is how
far the inverse-noise term decays before it is overtaken by the prior
term (see Fig.~\ref{fig:mask_solver}). In the case of analysing Planck
data, we find that the inverse-noise term decays by a factor of
roughly $\lambda=0.15$ at this point, which is unproblematic. In
simulations with lower resolution or higher noise, such that
$\lambda=0.01$, convergence is hurt substantially, and at
$\lambda=0.001$ the preconditioner breaks down entirely.

Since our own use-cases (which are targeted towards Planck) are not
affected by this restriction we have not investigated this issue very
closely.  We have however diagnosed the effect at low resolution with
a dense system solver for $(\Z \S^{-1} \Z)^{-1}$. In these studies, we
find that the problem is intimately connected with how the two
preconditioners are combined, and it may well be that more
sophisticated methods for combining preconditioners will do a better
job. For the interested reader, we recommend \cite{tang:2009} for an
introduction to the problem, as they cover many related methods
arising from different fields using a common terminology and
notation. In particular it would be interesting to use deflation
methods to ``deflate'' the mask subspace out of the solver.

Finally, we give a word of warning: In the full sky case, we have been able to rescale the CR
system arbitrarily without affecting the essentials of the system.
For instance, \cite{joint-bayesian-compsep} scales $\A$ with $\S^{1/2}$
so that the system matrix becomes
$
\I + \S^{1/2} \P^T \N^{-1} \P \S^{1/2}
$.
This has no real effect on the spherical harmonic preconditioners, but in pixel domain
it changes the shape of each term. The natural unscaled form of $\A$
is localized in pixel domain: The inverse-noise term essentially
looks like a sum of the instrumental beams, while the prior defines smoothness couplings
between a pixel and its immediate neighbourhood. However, $\S^{1/2}$ is a highly non-local operator,
and multiplying with this factor decreases locality and
causes break-down of our method. There may of course be other filters which would
increase locality in pixel domain instead of decrease it, in which case it could
be beneficial to apply them.

\subsection{Including a low-pass filter in the mask-restriction}

\newcommand{\find}{\quad}
\newcommand{\ind}{\quad \;}
\begin{figure*}
  \begin{minipage}{.48\linewidth}
    \small
    \begin{tabular}{l}
      $\text{Basic-V-cycle}(h, \b)$: \\
      \find {\bf Inputs:} \\
      \find  \ind $h$ -- The current level \\
      \find  \ind $\b$ -- Right-hand side \\
      \find \\
      \find \ind $H$ denotes the coarser level relative to $h$.\\
      \find {\bf Output:} \\
      \find \ind Approximation of $\G_h^{-1} \b$ \\
      \\
    \end{tabular}
    \begin{tabular}{ll}
      \find {\bf if} $h$ is bottom level: & \\
      \find \ind $\x \leftarrow \G^{+}_h \b$ & {\em By SVD pseudo-inverse} \\
      \find {\bf else}: \\
      \find \ind $\x \leftarrow \M_h \b$ & {\em Pre-smoothing} \\
      \find \ind $\r_h \leftarrow \b - \G_h\x$                & {\em Compute residual} \\
      \find \ind $\r_H \leftarrow \I_h^H \r_h$                & {\em Restrict residual} \\
      \find \ind $\ve{c}_H \leftarrow \text{Basic-V-Cycle}(H, \r_H)$ & {\em Recurse for coarse correction }\\
      \find \ind $\ve{c}_h \leftarrow (\I_h^H)^T \ve{c}_H$ & {\em Interpolate correction}\\
      \find \ind $\x \leftarrow \x +  \ve{c}_h$ & {\em Add correction} \\
      \find \ind $\x \leftarrow \x + \M_h (\b - \G_h \x_)$ & {\em Post-smoothing} \\
      \find {\bf return} $\x$
    \end{tabular}
  \end{minipage}
  \begin{minipage}{.48\linewidth}
    \small
    \begin{tabular}{l}
      $\text{Optimized-V-cycle}(h, \b)$: \\
      \find {\bf Inputs:} \\
      \find  \ind $h$ -- The current level \\
      \find  \ind $\b$ -- Right-hand side \\
      \find \\
      \find \ind $H$ denotes the coarser level relative to $h$.\\
      \find {\bf Output:} \\
      \find \ind Approximation of $\G_h^{-1} \b$ \\
      \\
    \end{tabular}
    \begin{tabular}{ll}
      \find {\bf if} $h$ is bottom level: & \\
      \find \ind $\x \leftarrow \G_h^{+} \b$  & \\
      \find {\bf else}: \\
      \find \ind $\x \leftarrow \M_h \b$ & \\
      \find \ind $\u \leftarrow \Y_h^T \x$ & {SHT at $L_h$, $N_h$} \\
      \find \ind $\tilde{\r} \leftarrow \Y^T_h \W_h \b - \ve{J} \D_h \u$ & {SHT at $L_H$, $N_h$}\\
      \find \ind $\r_H \leftarrow \Y_H \R_H \tilde{\r}$ & {SHT at $L_H$, $N_H$}\\
      \find \ind $\ve{c}_H \leftarrow \text{Optimized-V-Cycle}(H, \r_H)$ & \\
      \find \ind $\tilde{\ve{c}} \leftarrow \R_H \Y^T_H \ve{c}_H$ & {SHT at $L_H$, $N_H$}\\
      \find \ind $\x \leftarrow \x + \W_h \Y_h \tilde{\ve{c}}$ &  {SHT at $L_H$, $N_h$} \\
      \find \ind $\x \leftarrow \x + \M_h( \b - \Y_h \D (\u + \ve{J} \tilde{\ve{c}}))$ &  {SHT at $L_h$, $N_h$} \\
      \find {\bf return} $\x$

    \end{tabular}
  \end{minipage}

  \caption{Pseudo-code for the MG V-cycle. The matrices involved
    are defined in the main text.
    {\em Basic-V-Cycle:} The clean textbook version, exposing the basic
    structure of the algorithm. The important feature of the
    algorithm is that the solution vector $\x$ is never transferred
    directly between levels. Instead, a residual $\r_H$ is computed,
    which takes the role as the right-hand side $\b$ on the coarser
    level. The coarse solution is a correction $\ve{c}_H$ which is
    then added to the solution vector $\x$.  The resulting full
    V-cycle is a symmetric linear operator which can be used as a
    preconditioner for CG.  We keep recursing until there are less
    than 1000 coefficients left, and then solve using a pseudo-inverse based on the
    SVD, $\G_h^{+} \approx \G_h^{-1}$, as $\G_h$ may, depending on the size of the mask,
    be singular due to $\D_h$ being truncated at $L_h$.
    {\em Optimized-V-Cycle:} In this code we have
    inserted $\I_h^H = \Y_H \R_H \Y_h^T \W_h$ and $\G_h = \Y_h \D_h \Y_h^T$,
    and then reorganized the expressions so that the restriction and interpolation steps
    each share one SHT with the corresponding application of $\G_h$.
    The $\ve{J}$ operator denotes
    $\Y_h^T \W_h \Y_h$. The effect of this operator is to zero out any
    contribution that falls outside of the mask in the (full sky)
    spherical harmonic vectors; but numerical experiments indicate
    that the term can in practice be neglected also when using a small mask.
    In our numerical experiments we approximate $\ve{J} \approx \I$,
    reducing the total number of SHTs to 6 per level.
    The comments indicate the required resolution for each SHT, with $L_h$/$L_H$
    referring to fine/coarse harmonic band-limit, and $N_h$/$N_H$
    referring to fine/coarse grid.
  }
  \label{code:mg}
\end{figure*}

The feature that most strongly define the mask-restricted system
$\Z \S^{-1} \Z^T$ is the shape of the mask, and thus pixel
basis is the natural domain in which to approach this system.
The operator $\Z \S^{-1} \Z^T$ acts as
a convolution with an azimuthally symmetric kernel on the
pixels within the mask.  In Fig.~\ref{fig:mask_solver} we
plot a cut through this convolution kernel (blue in bottom panel).
Mainly due to the sharp truncation at the band-limit $L$, oscillations
extend far away from the center of the convolution kernel. To make the
system easier to solve we follow \cite{multigrid} and insert
a low-pass filter as part of the restriction operator $\Z$, so that
the projection from spherical harmonics to the pixels within the mask
is preceded by multiplication with the transfer function
\begin{equation}
  \label{eq:rl}
  r_\ell = \text{exp}(-\beta \ell^2 (\ell+1)^2),
\end{equation}
where we choose $\beta$ so that $r^2_{L/2}=0.05$. The
resulting system now has a transfer function of $r_\ell^2 / C_\ell$,
whose associated convolution kernel is much more localized  (dashed orange),
making it easier to develop a good solver for $\Z \S^{-1} \Z^T$.

After introducting this low-pass filter we no longer have
equality in Eq.~\eqref{eq:M_mask}, but only approximately that
$\Z \A \Z^T \approx \Z \S^{-1} \Z^T$.  This appears to not hurt the overall
method, as $r_\ell$ is rather narrow when seen as a pixel-domain convolution
(unlike $\sqrt{C_\ell}$, as noted above). Also note that we have now
over-pixelized the system $\Z \S^{-1} \Z$, as
higher-frequency information has been suppressed and the core of the
convolution kernel is supported by two pixels. Our attempts at
representing $\Z \S^{-1} \Z$ on a coarser grid failed however, because
the solution will not converge along the edge of the mask unless
the grid of $\Z$ exactly matches the grid of the mixing map $\Q_{\nu,k}$.
While the resulting system $\Z \S^{-1} \Z^T$ on the full-resolution grid
is poorly conditioned for the smallest scales, this does not prevent
us from applying iterative methods to solve for the larger scales.

\subsection{Multi-grid solver for the mask-restricted system}
\label{sec:multigrid}
Finally we turn our attention to constructing an approximate inverse
for $\Z \S^{-1} \Z^T$. We now write the same system matrix as
\begin{equation}
  \label{eq:W}
  \G = \Z \S^{-1} \Z^T = \Y \D \Y^T,
\end{equation}
where $\D$ is a diagonal matrix with $d_\ell = r_\ell^2 / C_\ell$
on the diagonal, and it should be understood that the spherical
harmonic synthesis $\Y$ only projects to grid points within the mask.

As noted in \cite{multigrid}, in the case where
$d_\ell \propto \ell^2$ this is simply the Laplacian partial differential equation
on the sphere, and the multi-grid techniques commonly used for solving
this system are also effective in our case.
We will focus on the case where $C_\ell$ is the CMB
power spectrum; in this case $1/C_\ell$ starts out proportional to
$\ell^2$, increasing to $\ell^6$ around $\ell \sim 1600$, eventually
reaching $\ell^8$ at $\ell \sim 6000$.  In theory this should make
the system harder to solve than the Laplacian, but it seems that in our
solver the application of the low-pass filter described in the previous
section is able to work around this problem.

To solve the system $\G \x = \b$ using iterative methods we might start out with a simple
diagonal approximate inverse,
\[
\M \equiv \text{diag}(\G)^{-1},
\]
which is in fact a constant scaling since $\G$ is locationally invariant. This turns out to work well as a
preconditioner for the intermediate scales of the solution.  For
smaller scales (higher $\ell$) the quickly decaying restriction
$r_\ell$ starts to dominate over $1/C_\ell$ such that the combined
effect is that of a low-pass filter; such filters can not to our knowledge be
efficiently deconvolved in pixel domain and an harmonic-domain preconditioner
would be required. Luckily, we do not need to
solve for these smaller scales, as the pseudo-inverse preconditioner
will find the correct solution in this regime, and the restriction operator $\Z$
will at any rate filter out whatever contribution comes from the solution of $\G$.

The problem at larger scales (lower $\ell$) is that the approximate
inverse would have to embed inversion of the coupling of two distant
pixels through a series of intermediate pixels in-between; this is
beyond the reach of our simple diagonal preconditioner.  For this
reason we introduce a \emph{multi-grid (MG) V-cycle}, where we
recursively solve the system on coarser resolutions. For each
coarsening, the preconditioner is able to see further on the sphere,
as indirect couplings in the full-resolution system are turned into
direct couplings in the coarser versions of the system. For a basic
introduction to multi-grid methods see e.g.  \cite{hackbusch:1985} or
the overview given in \cite{multigrid}.

The first ingredient in MG is an hierarchy of grids, which are denoted
relatively, with $h$ denoting an arbitrary level and $H$ denoting the
grid on the next coarser level.  We have opted for a HEALPix grid for $\Q_{\nu,k}$ and $\G$, and
use its hierarchical structure to define the coarser grid, simply
letting $N_{\text{side},H} = N_{\text{side},h} / 2$. We also need to consider which
subset of grid points to include to represent the region within the mask.
We got best results by only including those pixels of $H$ which are
covered 100\% by the mask in the fine grid $h$, so that no pixel on any level
ever represents a region outside the full-resolution mask.

The second ingredient in MG is the \emph{restriction operator} $\I_h^H$ which transfers
a vector from grid $h$ to grid $H$. We tried restriction operators
both in pixel domain and spherical harmonic domain, and spherical harmonic restriction
performed better by far. Thus we define
\begin{equation}
  \I_h^H = \Y_H \R_H \Y^T_h \W_h,
\end{equation}
where we use subscripts to indicate the grid of each operator,
and where $\R_H$ has some harmonic low-pass filter $r_{H,\ell}$
on its diagonal.
In \cite{multigrid} the corresponding filter had to be carefully
tuned to avoid problems with ringing, because the  $\N^{-1}$-term created
high contrasts in the system matrix. In the present method we no longer have
to deal with the $\N^{-1}$ term, and the requirements on the low-pass filter
are much less severe, as long as they correspond to a convolution
kernel with a FWHM of roughly one pixel on the coarse grid. A Gaussian band-limited at $L_H=3 N_\text{side,H}=L_h/2$
performed slightly better than the filter of Eq.~\eqref{eq:rl} in our tests, even if
it has somewhat more ringing at this band-limit.

The third ingredient in MG is the coarsened linear system.
\begin{equation}
\G_H = \I_h^H \G_h \I_H^h = \Y_H \D_H \Y_H^T,
\end{equation}
where $d_{H,\ell} = r^2_{H,\ell} d_{h,\ell}$ is band-limited at
$L_H = L_h/2$\footnote{Note that the matrix coarsening must be done in another way if
  using a pixel-domain restriction operator. In that case
  $\D_H$ is in principle a dense matrix due to pixelization
  irregularities, but can still be very well approximated by a diagonal
  matrix. Details are given in Appendix~\ref{appendix:mask-solver}.}.
We stress again that the grid $H$ embeds the structure of the mask,
so that $\Y_H$ in this context denotes spherical harmonic synthesis only to grid
points within the mask. In computer code, zero padding is used outside of the mask before
applying $\Y_H^T$, and entries outside the mask are discarded after applying $\Y_H$.

The fourth ingredient in MG is an approximate inverse, in this context named the \emph{smoother}.
The name refers to removing small scales in the error $\x_\text{true} - \x$.
Removing these scales happens through approximately solving the system, and should
not be confused with applying a low-pass filter.
In our case we will use the simple constant smoother $\M$ discussed above,
although in combination with a damping factor $\omega = 0.2$, so that
the eigenvalues of $\omega \M \G$ are bounded above by 2 as required
by the MG method. We write $\M_h = \omega \; \text{diag}(\G_h)^{-1}$
for the smoother on level $h$.

Finally, the ingredients are combined in the simplest possible MG V-cycle
algorithm (see Fig.~\ref{code:mg}). It turns out that the restriction and interpolation
operations can share one SHT each with the associated system matrix multiplication,
so that 6 SHTs are required per level. Furthermore, the SHTs can be performed
at different resolutions for additional savings.

\begin{figure}
  \centering
  \includegraphics[width=1\linewidth]{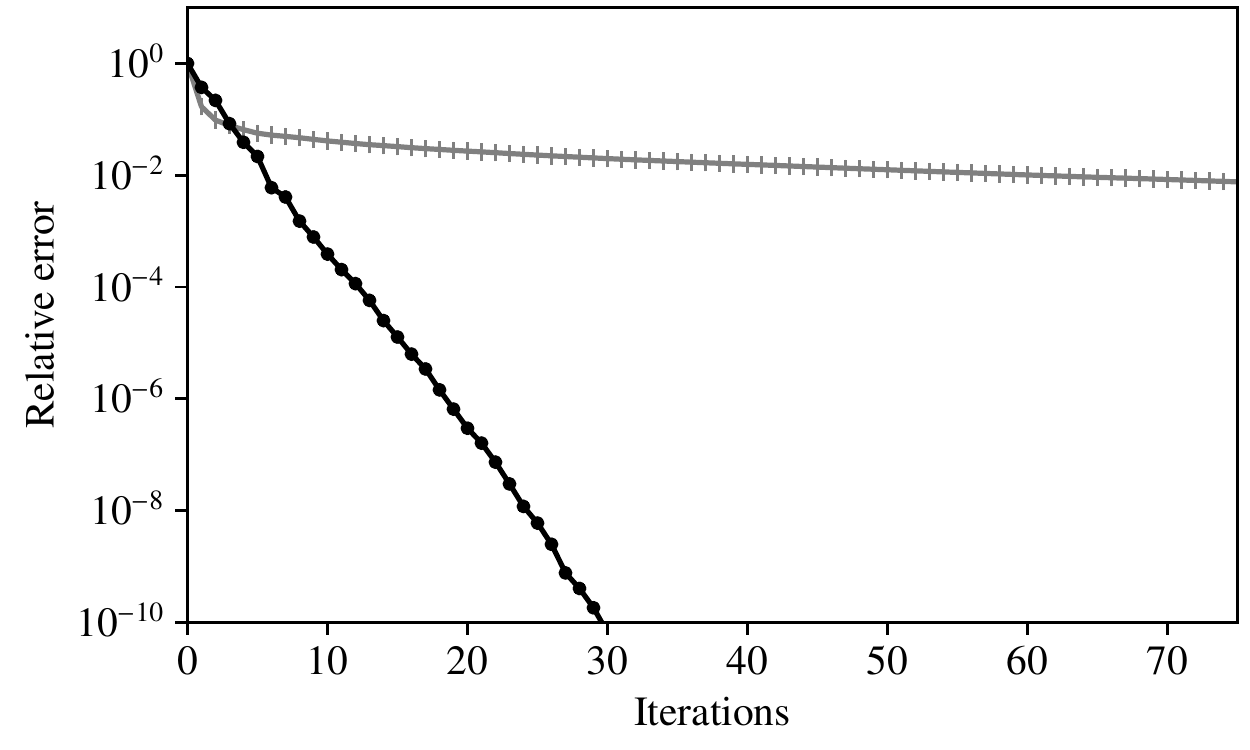}
  \includegraphics[width=.7\linewidth]{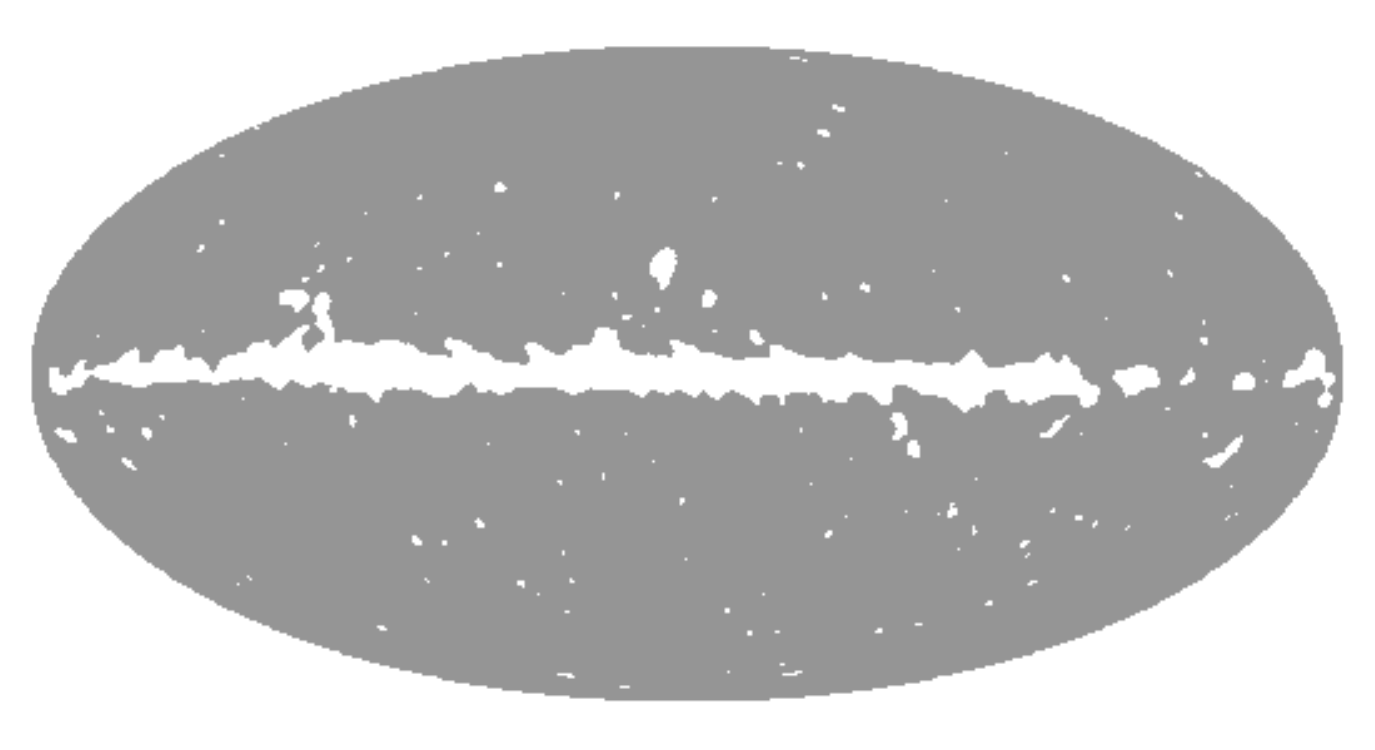}
  \caption{Convergence of the $\M_\text{add}$ preconditioner (top
    panel) when including a sky mask in the model (bottom panel). We
    fit a single CMB component to hypothetical foreground-cleaned maps
    on all 9 Planck bands, specifying a fiducial $C_\ell$ prior for
    the CMB power spectrum.  We plot the convergence when using
    a diagonal preconditioner (gray ticks) and the $\M_  \text{add}$ preconditioner
    developed in Sect.~\ref{sec:mask} (black circles).
    See Sect.~\ref{sec:benchmarks} for details on the benchmark setup.
  }
  \label{fig:benchmark_mask}
\end{figure}

Figure~\ref{fig:benchmark_mask} shows the results of solving the full system
when inserting this algorithm as an approximation
of $(\Z \S^{-1} \Z^T)^{-1}$ in Eq.~\eqref{eq:M_mask}.
While the diagonal preconditioner degrades, $\M_\text{add}$
converges very quickly. The pseudo-inverse preconditioner $\M_\text{PI}$ by itself shows
much the same behaviour as the diagonal preconditioner in this
situation (not plotted).


\section{Benchmark notes}
\label{sec:benchmarks}
The implementation used for the convergence plots in this paper are produced using
a prototype implementation written in a mixture of Fortran,
Cython and Python, and is available under an open source
license at \url{https://github.com/dagss/cmbcr}.
As a prototype, it does not support polarization or distributed computing with MPI.
A full implementation in the production quality Commander code is in progress.

\begin{figure}
  \centering
  \includegraphics[width=1\linewidth]{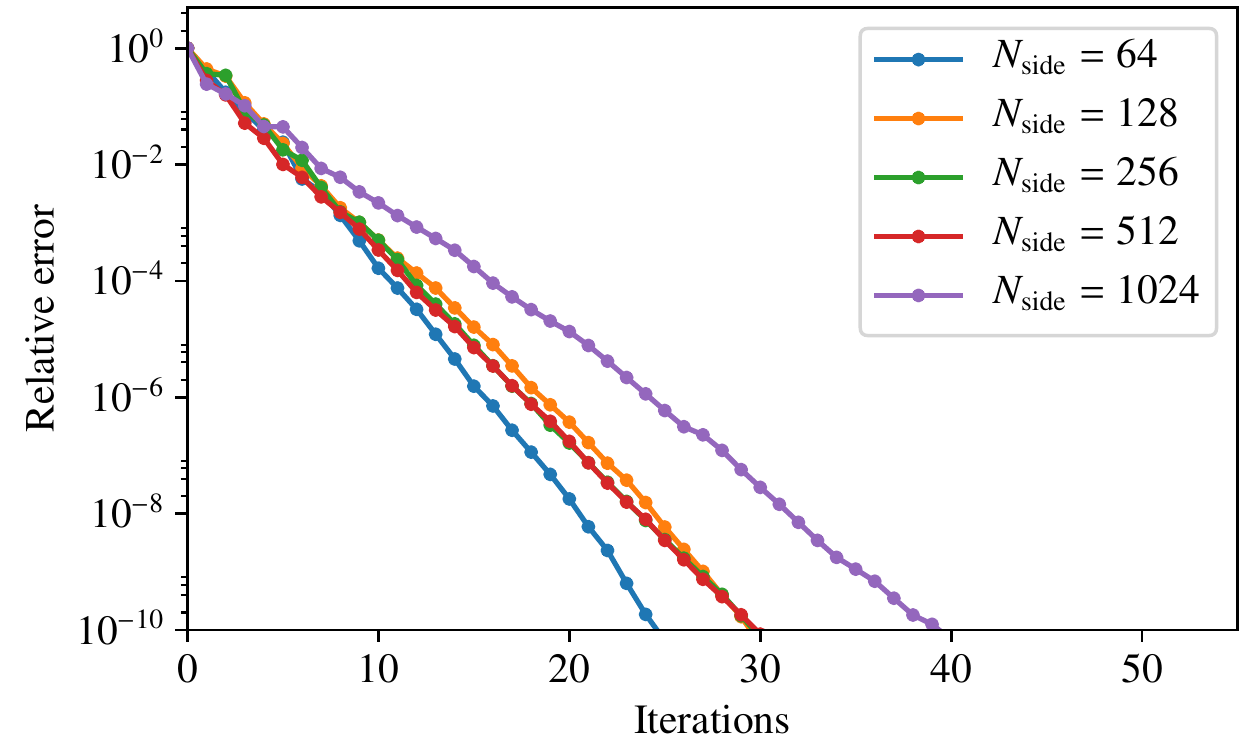}
  \caption{Convergence rate of the solver at different resolutions,
    using the setup with a single-component model and a mask
    corresponding to Fig. \ref{fig:benchmark_mask}.  While the
    iteration count varies depending on resolution, other
    model parameters, such as the size and shape of the mask, the
    prior, and the signal-to-noise ratio, has an equally strong effect
    on convergence speed, and we believe the algorithm to be resolution-neutral
    on a fundamental level due to the multi-grid nature. The resolution parameter $\Nside$ here
    refers to the resolution of the inverse-noise map.  The band-limit
    $L$ is then scaled down by a factor of $\Nside/2048$, from an
    assumed full-model band-limit of $L=6000$.  }
 \label{fig:benchmark_compare_resolutions}
\end{figure}

As the solver does not support MPI-parallelization,
a full resolution run is an overnight run.
However, as shown in Fig. \ref{fig:benchmark_compare_resolutions}, the performance of
our solver does not significantly degrade as resolution is increased, and in this paper we
therefore present benchmarks of down-scaled systems that we have worked with during
development. The downgrade procedure follows these steps:
\begin{itemize}
\item Downgrade the RMS maps to a lower $\Nside$ using HEALPix routines
\item Find the best fit Gaussian beam approximation to the instrumental beams, and
  make equivalent low-resolution beams based on scaling down the FWHM parameter
\item Downgrade each prior $C_\ell$ by sub-sampling coefficients. For instance, for a degrade
  from $\Nside=2048$ to $\Nside=256$, we take every 8th coefficient. Similarly,
  we divide each band-limit $L_k$ by the relevant downgrade factor.
\item Scale $C_\ell$ in such a way that the diagonal of $\S^{-1}$ crosses
  the diagonal of $\P^T \N^{-1} \P$ at the same $\ell$, relative to the full $L_k$,
  ensuring that the system has the same signal-to-noise properties as the full resolution system.
\end{itemize}
To save computing resources, the numerical
experiments of Figs.~\ref{fig:benchmark_single},
\ref{fig:benchmark_compsep} and \ref{fig:benchmark_mask} are performed
on system downgraded to $\Nside=128$.

Simulations are performed with a known $\x_\text{true}$ drawn randomly from a Gaussian distribution,
and a right hand side given by $\b = \A \x_\text{true}$. Then the convergence statistic denoted ``relative
error'' in these figures simply reads
\[
e_i \equiv \|\x_i - \x_\text{true} \|.
\]
Finally, unless otherwise noted, we add regularization noise to the
1\% of the highest signal-to-noise pixels in the RMS maps. As noted in
Fig.~\ref{fig:benchmark_single}, this is more of an advantage for the
diagonal preconditioner than the pseudo-inverse preconditioner, but
this typically mimics what one would do in real analysis cases.

In the present paper we have focused strictly on algorithm
development, and as such the prototype code is not optimized; we
have not invested the effort to benchmark the preconditioners in terms
of CPU time spent. As detailed in Sects.~\ref{sec:full-sky-multi-comp}
and \ref{sec:multigrid}, the additional cost for a particular use-case
can be calculated from the number of extra SHTs.

The block-diagonal preconditioner we use as a comparison point is described in further
detail by \cite{gibbs-eriksen}. In the notation of this paper, it can be written
\[
\M_\text{diag} \equiv (\U^T \text{diag}(\T) \U)^{-1},
\]
where each element of $\text{diag}(\T)$ can be computed in $O(L)$ time
by a combination of Fourier transforms and computing the associated
Legendre polynomials, which is available, e.g., in the
latest version of Libsharp\footnote{\url{https://github.com/dagss/libsharp}}.
Since the matrix $\U^T \text{diag}(\T)\U$ consists of $(\lmax + 1)^2$ blocks of
size $\Ncomp \times \Ncomp$ the inversion is cheap.

\section{Discussion and outlook}
\label{sec:conclusions}

In this paper we have presented a versatile Bayesian model for the
multi-resolution CMB component separation and constrained realization
problem, as well as an efficient solver for the associated linear
system.  This model is currently in active use for component
separation for the Planck 2017 collaboration. The final result is the
ability to perform exact, full-resolution, multi-resolution component
separation of full-sky microwave data within a reasonable number of
conjugate gradient iterations.

To achieve such good convergence several novel techniques were
employed. First, we developed a novel pseudo-inverse based
preconditioner. For the full-sky case this provides a speed-up of
2--3x compared to a diagonal preconditioner, depending on model
parameters.  Second, we extend the model with a mask through the
mixing maps, rather than through the noise covariance matrix, to avoid
ringing problems associated with going between spherical harmonic
domain and pixel domain.  Third, we solved for the solution under
the mask using a dedicated multi-grid solver in pixel domain
restricted to the area under mask, where the linear system can be
simplified.

We note that the pseudo-inverse preconditioner not only performs very
well for the full-sky case with reasonably uniform mixing maps, but it
is also very simple to implement; significantly simpler than the
previously standard diagonal preconditioner. As mentioned earlier,
this technique is of course not novel for or restricted to CMB
component separation; it has been in use for solving Navier-Stokes
equations for some time. The fundamental idea is to approximate the
total inverse of a sum with the best linear combination of inverses of
each term.  A problem related to this paper, which in particular fits
this description, is the basic CMB map-making equation
\citep[e.g.,][]{tegmark1997}. This equation is a sum over individual
time segments of observations, which can in isolation be inverted in
Fourier domain. If the pseudo-inverse preconditioner works well in
this case, as we believe it will, it may speed up exact maximum
likelihood map makers substantially.


Regarding more direct extensions of the work in the present paper, a
natural next step is the use of some pixel domain basis instead of a
spherical harmonic coefficients to represent the microwave
components. We have so far have assumed an isotropic prior for all
components which can be specified in the form of a power spectrum
$C_\ell$, with a sharp band-limit $L$. This model has a tendency to
excite ringing in the resulting maps around sharp objects, unless much
time is spent tuning the priors, or one adopts a very high band-limit
$L$ for all components. Working with pixel-domain vectors and, with
the exception of the CMB, pixel-domain prior specifications, one could
easily define the models that more robust against this problem.  Also,
we know that the diffuse foregrounds has much variation where their
signal is strong, but should be more heavily stabilized where their
signal is weak. Such a non-isotropic prior is easier to model using a
sparse matrix in pixel domain. Of particular interest are the
so-called Conditional Auto-Regressive (CAR) models, which have a
natural interpretation and which directly produce sparse inverses
$\S^{-1}$.

\begin{acknowledgements}
  We thank Jeff Jewell, Sigurd Næss, Martin Reinecke, and Mikolaj
  Szydlarski for useful input. Parts of this work was supported by the
  European Research Council grant StG2010-257080. Some of the results
  in this paper have been derived using the HEALPix \citep{healpix},
  Libsharp \citep{libsharp}, Healpy, NumPy, and SciPy software packages.
\end{acknowledgements}

\bibliographystyle{aa} \bibliography{pseudoinv}

\begin{appendix}

\section{Details of the pseudo-inverse preconditioner}
\label{appendix:pseudo-inverse}
\subsection{Approximating the inverse-noise maps}
\label{appendix:approximating-inverse-noise-map}
We seek $\alpha_\nu$ such that the distance between
$\widetilde{\N}_\nu^{-1}$ and the identity matrix is minimized:
\begin{align*}
  \| \widetilde{\N}_\nu^{-1}(\alpha_\nu) - \I \|_2
  &= \| \alpha^{-2}_\nu \Y^T_\nu \N^{-1}_\nu \Y_\nu - \Y_\nu^T \W_\nu \Y_\nu \|_2  \\
  &= \| \Y^T_\nu ( \alpha^{-2}_\nu \N^{-1}_\nu - \W_\nu )\Y_\nu \|_2  \\
  &= \| \Y^T_\nu \W^{1/2} ( \alpha^{-2}_\nu \W^{-1} \N^{-1}_\nu - \I ) \W^{1/2} \Y_\nu \|_2   \\
  &=  \| ( \alpha^{-2}_\nu \W^{-1} \N^{-1}_\nu - \I ) \|_2.
\end{align*}
The last equality follows because all the singular values of $\Y^T
\W^{1/2}$ are 1, at least for the Gauss-Legendre grid. For the HEALPix
grid the statement is only approximate, within 10-20\%, depending on
resolution parameters, and this is close enough for our purposes.  We
conclude that the best choice is
\begin{equation}
  \label{eq:optimal-alpha}
\alpha_\nu = \sqrt{\frac{\sum_i (\tau_\nu(\hat{n}_i)/w_i)^2}{\sum_i \tau_{\nu}(\hat{n}_i)/w_i}},
\end{equation}
where $\tau_i$ represents the pixels in the inverse-noise variance map
on the diagonal of $\N^{-1}_\nu$, and $w_i$ are the quadrature weights
of the associated grid.  We have verified this expression
experimentally by perturbing $\alpha_\nu$ in either direction, and
find that either choice leads to slower convergence. Ultimately, the
method may even fail to converge if $\alpha_\nu$ deviates too much
from the optimal value.

In code, the easiest way to compute $\tau_\nu(\hat{n}_i)/w_i$ is by
performing a pair of SHTs, $\W^{-1} \tau = \Y \Y^T \tau$.  By
replacing the usual analysis $\Y^T \W$ with adjoint synthesis $\Y^T$,
we end up implicitly multiplying $\tau$ with $\W^{-1}$.

\subsection{Approximating the mixing maps}
\label{appendix:approximating-mixing-map}
Following a similar derivation to the previous section, the optimal scalar
to approximate the mixing maps is given by minimizing
\[
\|\WQ_{\nu,k} - q_{\nu,k} \I \|
= \| \Y^T \W \Q_{\nu,k} \Y - q_{\nu,k} \Y^T \W \Y \|
= \|\Q_{\nu,k} - q_{\nu,k} \I \|,
\]
so that the best choice is
\begin{equation}
  \label{eq:optimal-q}
q_{\nu,k} = \frac{\sum_i q_{\nu,k}(\hat{n}_i)^2}{\sum_i q_{\nu,k}(\hat{n}_i)}.
\end{equation}
In our cases, however, the difference between this quantity and the
mean of the mixing map is negligible.

\subsection{Possible future extension: Merging observations}
\label{sec:future-extensions}
Depending on the data model, it may be possible to reduce the number
of SHTs required for each application of the pseudo-inverse
preconditioner.  When $\Nobs > \Ncomp$, the system in some ways supply
redundant information.  Assume that two rows in $\U$ are (at least
approximately) identical up to a constant scaling factor; this
requires that the corresponding sky maps have the same beams, the same
normalized spatial inverse-noise structure, and is located at the same
frequency $\nu$ with the same SED for each component.  That is, we
require both $\widetilde{\N}_1^{-1} = \widetilde{\N}_2^{-1}$ and $\U_2
= \gamma \U_1$, where $\U_\nu$ indicate a row in $\U$ and $\gamma$ is
an arbitrary scale factor. Note that this situation is very typical
for experiments with several independent detectors within the same
frequency channel, which is nearly always the case for modern
experiments.

Under these assumptions, we have
\begin{gather*}
\mat{ \U_1^T &  \U_2^T}
\mat{ \N_1^{-1} & \\ & \N_2^{-1} }
\mat{ \U_1 \\ \U_2} =
\\
\mat{ (\gamma + 1)^{1/2} \U_1^T }
\mat{ \N_1^{-1} }
\mat{ (\gamma + 1)^{1/2} \U_1 }.
\end{gather*}
Thus, we may combine the two rows without affecting the rest of the
system, and thereby halve the number of SHTs required. Of course, two
sky observations with such identical properties could have been
co-added prior to solving the system, as is typically done when
creating co-added frequency maps. In practice, however, there are
typically many advantages in working with detector sub-sets, including
improved ability to isolate systematics effects
\citep[e.g.,][]{planck2015:foregrounds}, and more easily allow for
cross-correlation analysis. In addition, there may be experiments
where some sky maps do not have identical properties and one does not
wish to co-add, but they are similar enough that co-adding poses no
problem if done in the preconditioner alone. One then needs to somehow
produce compromises for $\B$, $\N^{-1}$ and $\M_{\nu,k}$, replace the
relevant matrices with the compromise versions, and finally use the
row merge procedure described above to create a new $\U$ solely for
use in the preconditioner.

\section{Alternative strategies for the mask-restricted solver}
\label{appendix:mask-solver}

We spent some time
exploring pixel-domain restriction operator before turning to spherical
harmonic restrictions.  The simple restriction we
attempted, averaging the 4 nested pixels using the standard HEALPix
\verb@udgrade@ function, more than doubled the number of iterations required
for a small mask when compared to a restriction in spherical harmonic domain,
and had trouble converging at all for a large mask.
The spherical harmonic restrictions are therefore well worth the extra time
spent performing SHTs. Still, it is probably possible to pull out a little bit
more performance by experimenting with averaging over a larger region
with a choice of weights that approximates a Gaussian
low-pass filter.

When using a pixel based restriction the system can no longer be coarsened
simply by multiplying spherical harmonic transfer functions. However,
since the operator is rotationally and
locationally invariant it is simple to coarsen the system numerically.
The idea is to image the operator
in a single pixel, and then solve for the spherical harmonic transfer
function that would produce this image. Let $\u$ represent a unit
vector located on equator on the coarse grid $H$. We then seek $\D_H$
such that
\begin{align*}
\Y_H \D_H \Y^T_H \u &= \I_h^H \G_h \I_H^h \u = \I_h^H \Y_h \D_h \Y_h^T \I_H^h \u \\
\D_H \Y^T_H \u &= \Y_H^T \W_H \I_h^H \Y_h \D_h \Y_h^T \I_H^h \u.
\end{align*}
Now, assuming that $\D_H$ is diagonal we must have
\begin{align*}
  d_{H,\ell,m} = \frac{(\Y_H^T \W_H \I_h^H \Y_h \D_h \Y_h^T \I_H^h \u)_{\ell,m}}{(\Y^T_H \u)_{\ell,m}}.
\end{align*}
In practice, due to pixelization effects, $\D_H$ cannot be fully
diagonal and this equation cannot be satisfied for all $\ell$,
$m$. However, assuming that $\D_h$ is isotropic it should be fully
characterized by the modes $m=0$, and, as $\u$ was located on equator,
these produce a very good estimate.  Using this coarsening procedure
instead of the analytical coarsening procedure in
Sect.~\ref{sec:multigrid} produces identical results when applied to
the Gaussian restriction operator.  With pixel-domain restriction
operators, pixelization effects will hurt the approximation somewhat.
We expect that the approximation will be hurt less if the averaging weights
are a function of the physical distance between the grid points rather
than the logical distance.

We have also experimented with using Fourier basis to represent $\G_h$.
When using a thin mask around equator, or a small point source, applying
torus- or flat sky-approximations, respectively,
allows for using the much faster FFTs instead of SHTs.
The operator should then be transferred using the same principle
as above. Let $\F$ denote a discrete Fourier transform from harmonic space to real space,
then, within a narrow equatorial band or a small patch, we require
\begin{equation}
  \label{eq:fft}
  \F \D_\text{FFT} \F^T \u \approx  \Y \D_\text{SHT} \Y \u
\end{equation}
and solve for
\[
  \D_{\text{FFT},k,k'} = \frac{(\F^{-1} \Y \D_\text{SHT} \Y^T \u)_{k,k'}}{(\F^T \u)_{k,k'}}.
\]
Then coarser systems can be produced either analytically (restriction
in harmonic domain) or by appropriate modifications to the technique above
(restriction in pixel domain).
We were able to produce a functional solver using this principle, but
feel that the loss in flexibility was not worth the gain in performance
compared to the solver presented in Sect.~\ref{sec:multigrid}.

\end{appendix}

\end{document}